\documentclass[a4paper,12pt]{article}
\usepackage{amsmath}
\usepackage{amsthm}
\usepackage{amsfonts}
\usepackage{amssymb}
\usepackage{mathrsfs}
\usepackage{graphicx}
\usepackage{color}
\usepackage[colorlinks=true, allcolors=blue]{hyperref}

\newtheorem{theorem}{Theorem}
\newtheorem{lemma}{Lemma}
\newtheorem{remark}{Remark}

\newtheorem{corollary}{Corollary}
\newtheorem{proposition}{Proposition}

\usepackage[english]{babel}

\numberwithin{equation}{section}

\hypersetup{ pdftitle={Draft},
pdfauthor={Emir S. Fadhilla, Bobby E. Gunara}, 
pdfkeywords={Local Esistence, Dyonic Black Hole}, 
bookmarksnumbered, pdfstartview={FitH}, urlcolor=blue, }

\begin{document}
\pagestyle{plain}




\title{\LARGE\textbf{Global Existence and Singularity Formation of Classical Solutions in Einstein-Skyrme System}}

\author{Emir Syahreza Fadhilla$^{\sharp,\ddagger}$, 
Ardian Nata Atmaja$^{\ddagger}$, and Bobby Eka Gunara$^{\sharp, \ddagger}$\footnote{Corresponding author}\\ \\
$^{\sharp}$ \textit{\small Theoretical High Energy Physics Research Division, }\\
\textit{\small Faculty of Mathematics and Natural Sciences,}\\
\textit{\small Institut Teknologi Bandung}\\
\textit{\small Jl. Ganesha no. 10 Bandung, Indonesia, 40132}
\\ {\small and} \\
$^{\ddagger}$\textit{\small Research Center for Quantum Physics, National Research and Innovation Agency (BRIN),}\\
\textit{\small Kompleks PUSPIPTEK Serpong, Tangerang 15310, Indonesia}\\
\\
\small email: emirsyahreza@students.itb.ac.id, ardi002@brin.go.id, bobby@itb.ac.id}

\date{\today}

\maketitle




\begin{abstract}

In this paper, we elucidate the problem of gravitating Skyrmion governed by field equations of the Einstein-Skyrme system with no potential term in the Bondi coordinate. The spherical symmetry has to be assumed and both the metric functions and Skyrme ansatz depend on radial and retarded time coordinates which implies that the system is dynamic. We show that unique smooth solutions with arbitrary initial data exist for a certain time interval with the extra condition that the time interval must be finite in order to have a non-zero topological charge. Then, the strategy to show that global smooth solutions exist is to restrict the initial data and proceed to find suitable configurations for extending the time interval to infinity. We also discuss the possible configurations within The Einstein-Skyrme System which develop singularities in coordinate origin.
\end{abstract}




\section{Introduction}
\label{sec:intro}

The Skyrme model was initially proposed as a model for strongly interacting particles in flat spacetime and is proven to be a good model in low energy region. It is constructed from the usual sigma model with a \(S^3\) target space and an extra quartic term, instead of just the usual quadratic term, in the Lagrangian density of the model. The quartic term is formulated in such a way that the resulting term is still proportional to the square of time derivative of the field. Furthermore, this model possess a conserved current known as topological current that is proportional to the volume form on the target space. The corresponding charge is considered as the baryon number and the model is considered as the unified field theory of mesons and baryons \cite{Skyrme:1961vq,Skyrme:1962vh}. The classical solution of the dynamical equation in Skyrme model are known as Skyrmion, which belong to the topological solitons due to the topological properties that have been mentioned above and it propagates in the form of a solitary wave that does not obey superposition principles due to its non-linear nature \cite{Manton:2004tk}.
The Einstein-Skyrme system is the system of dynamical field equations arising from the least-action principle of the following action
\begin{equation}\label{EinsteinHilbertAction}
    \mathcal{S}=\int\sqrt{-\gamma}\frac{R}{16\pi}~d^4x+\mathcal{S}_{\text{Skyrme}},
\end{equation}
where \(\gamma\) is the determinant of metric tensor, \(R\) is the Ricci scalar of  corresponding spacetime manifold and \(\mathcal{S}_{\text{Skyrme}}\) is the action of the Skyrme model in \(3+1\) dimensional spacetime \cite{Gunara:2021jvq}. The expression of the action of the Skyrme model is given by
\begin{eqnarray}
\mathcal{S}_{\text{Skyrme}}=-\int\sqrt{-\gamma}\left[C_1\phi^a_\kappa\phi^a_\tau \gamma^{\kappa\tau}+C_2\phi^a_{[\kappa}\phi^b_{\alpha]}\phi^a_{[\tau}\phi^b_{\beta]}\gamma^{\kappa\tau}\gamma^{\alpha\beta}\right]d^4x\label{SkyrmeAction}
\end{eqnarray}
where \(\phi^a\)'s are the scalar multiplets and \( C_1, C_2 \in \mathbb{R}^+\) are real coupling constants. As we know, the Euler-Lagrange equations corresponding to the variation in metric tensor components are equivalent to Einstein's field equations with the Skyrme field acting as its matter field. This system models the interaction between Skyrmion and gravity that are minimally coupled.

The study of Skyrmions under the effect of gravity is first introduced in \cite{Bizon:1992gb,Droz:1991cx} where it 
 shows that such a model admits a large family of spherically symmetric asymptotically flat solutions. The stable spherically symmetric solutions exist for both with and without horizons provided that the coupling constant is greater than a certain critical value \cite{Heusler:1991xx}. This led to an extensive study on the Skyrmion hairy black hole as a candidate for counter-example of the no-hair conjecture, for example in \cite{Luckock:1986tr} where the Skyrme field is located on a Schwarzschild background due to the assumption that the gravitational coupling is small \cite{Volkov:2016ehx}. The result proposes that it is possible for a black hole to have a Skyrmion hair, but this model was, actually, incomplete because the back-reaction from the Skyrme field to the gravity is ignored. Later, it is found that stable black holes with back-reacting massless Skyrmion hair exist for asymptotically AdS four-dimensional spacetime, reported in \cite{Shiiki:2005xn}. In this setup, the Skyrmion possess a fractional topological charge, in contrast with the usual Skyrmion in flat spacetime which has an integer-valued topological charge. This property is quite peculiar because it implies that the gravitational collapse of a Skyrme field to a black hole does not conserve the topological charge. This property is also found in a more general system where higher order terms are included \cite{Adam:2016vzf,Gudnason:2016kuu} and also in higher dimensional cases such as five-dimensional asymptotically flat Skyrme black hole \cite{Brihaye:2017wqa} and AdS black holes in higher dimensional Einstein-Skyrme system \cite{Gunara:2021jvq}. 

It is also possible to construct a star-like solution for a gravitating Skyrmion. The star solution does not possess a horizon and typically has an integer-valued topological charge. Some of the known solutions are gravitating BPS Skyrmions as a neutron star \cite{Adam:2014dqa,Naya:2019rlm,Adam:2020yfv}, and a more general five-dimensional Skyrmions and Skyrme stars \cite{Brihaye:2017wqa,Fadhilla:2020rig} which includes up to the octic terms of the Skyrme Lagrangian. All of the solutions from gravitating Skyrmion systems mentioned above have a similarity, namely, they assume time-independency by taking the ansatz to be either static or stationary. From the known examples, we conclude that the time dependency in those models is introduced perturbatively around the time-independent solutions to study the stability under small, time-dependent, disturbances. Thus, it is interesting to generalize the study of the Einstein-Skyrme system by assuming time-dependency on the solutions. Such solutions could become the model of Skyrmions dynamics under gravitational effect and show us how the static and stationary solutions can form in their steady-state limit.

It is known that the time-dependent solution of simpler gravitating scalar field systems do exist. The study is first proposed in \cite{Christodoulou:1986du,Christodoulou:1986zr} where the authors shows that unique spherically symmetric global solutions of Einstein-Klein-Gordon system do exist. Furthermore, they showed that such system could develop naked singularity in finite time \cite{Christodoulou:1994hg} although the singularity turns out to be unstable, hence it cannot be the counter-example of the cosmic censorship conjecture \cite{Christodoulou:1999math}. Regardless of the results found in the singularities in Einstein-Klein-Gordon system, the method demonstrated in those works shows promising future and can be developed further for a wider class of gravitating scalar fields. An example of this can be found in \cite{Chae:2001ec} where the author shows that spherically symmetric global solutions of the Einstein-Klein-Gordon system with a special type of polynomial potential exist for small initial data. It is also possible to adopt the method for gravitating scalar multi-fields where the fields possess internal symmetries. 

In this work, we propose that a similar method to show the global existence of the Einstein-Klein-Gordon system can be adopted for a more general Einstein-Skyrme system. The idea of the proof is as follows: The Skyrme model is characterized by a charge, known as the topological charge, and it coincides with the Klein-Gordon model when the topological charge, \(B\), is set to be zero. While the \(B=0\) limit of the model (the Klein-Gordon model) is a linear model, the Skyrme model in general, with \(B\geq0\), is a non-linear scalar theory which is characterized by the scalar fields equipped with internal symmetries, that is, the symmetries of the target space (see \cite{Manton:2004tk} for further details on non-linear scalar theory). 
This raises a challenge since we cannot solve the components of the Skyrme field multiplet independently due to the constraint that it must preserve the internal symmetry. A common solution to circumvent this complexity is the introduction of a coordinate system on the target space. As such, the coordinates on the target space act as the new dynamical fields which we can guarantee to obey the symmetries of its internal structure. Due to the non-linear nature of the theory, we choose to work with a certain ansatz in order to simplify the problem which serves as the coordinate system on the target space mentioned before. These ansatzes also obey the assumed symmetry of the spacetime and in this work, we choose to work with the dynamical hedgehog ansatz that is suitable for spherically symmetric models with time dependence. This way, we can reduce the complexity of the problem of non-linear scalar theory into a system of non-linear wave equations.
 
Furthermore, the dynamical equation of the Skyrme model itself is highly non-linear in the sense that we cannot superpose two independent solutions\footnote{This property is known as the common property of soliton. A solitary wave solution of non-linear dynamical equations.}.
This non-linearity does not just come from the implementation of ansatz, but also from the higher power derivative terms and potential terms in the action of the Skyrme model which is explicitly given in \eqref{SkyrmeAction}. Within the spherically symmetric hedgehog ansatz, the dynamical equations for the system are the set of Einstein's equations supplied by the conservation of energy-momentum tensor that is equivalent to the Euler-Lagrange equations of action \eqref{SkyrmeAction} for non-trivial solutions. In order to use Christodoulou's method, proposed in \cite{Christodoulou:1986zr}, to the Einstein-Skyrme system we need to find well-behaving estimates that are always bigger than the terms in the dynamical equations but decay fast enough at spatial infinity. We found that polynomials are able to estimate such non-linear terms in the Einstein-Skyrme system. This approach transforms our problem from highly non-linear wave equations into a slightly simpler problem, that is, non-linear wave equations with only polynomial terms. A similar problem of wave equations with polynomial terms has been addressed by Chae in \cite{Chae:2001ec} with Christodoulou's method, as mentioned previously. Thus, the main procedure that is frequently employed in this work is finding appropriate polynomial estimates for the terms in the wave equations of the Einstein-Skyrme system.

The idea for the proof of local existence is to find a sufficiently small finite time interval where we can guarantee that the smooth solutions are locally Lipschitz. This approach is similar to Christodoulou's approach to the Einstein-Klein-Gordon system, except that we employ the estimates and Lipschitz condition to the ansatz instead of the field, \(\phi\). Thus, the resulting estimates highly depend on the corresponding coordinate system and its assumed symmetries. The idea for the existence of globally regular solutions follows similar reasoning, that is, in order to find globally regular solutions (with semi-infinite time intervals) we constrain the norm of the initial data to be bounded above. The bound for the intervals of local solutions and the norm of the initial data for global solutions are deduced from Lipschitz's condition that must be satisfied to ensure uniqueness. On the other hand, the topology of the Skyrmions forbids the maximum values of initial data to be less than multiples of \(\pi\) \cite{Skyrme:1962vh, Manton:2004tk}. Thus, we expect that the initial data is both bounded from above and below.

It is also known in the soliton theory that such a non-linear system could develop singularities in finite time under certain conditions \cite{drazin_johnson_1989}. These singularities in the Einstein-Skyrme system are not directly evident in the case of travelling wave solutions considered within Christodoulou's method. To observe the possible singular solutions we consider the self-similar solutions of the Einstein-Skyrme system. In order to do this, firstly, we study the existence and uniqueness of smooth self-similar solutions in a semi-infinite interval of self-similar coordinates. From this step, we found that unique self-similar solutions do exist with a remark that in order to include the origin vertex of the coordinate system (\(r=0\)) the ansatz of the Skyrme field must satisfy the boundary condition of topological Skyrmion. The implication of this boundary condition is that some of the metric tensor components have to change their sign at a finite distance from the origin, implying the existence of a horizon. From here, we proceed to show that the self-similar solutions we found above are, indeed, singular at the origin after a finite interval of time. This result implies that the singularity is possibly hidden inside a horizon, in contrast with the result found for the Einstein-Klein-Gordon system where the singularity is naked \cite{Christodoulou:1994hg}.

The article is organized as follows: Section \ref{Sect2} is devoted to introduce the details of the Einstein-Skyrme system as a model of gravitating Skyrmions. In this section, we describe the characteristics of the Skyrme model and how we can formulate the system of equations under the assumption of spherical symmetry. In Section \ref{Sect3} we show the main results for both the local and global existence of spherically symmetric time-dependent solutions in the Einstein-Skyrme system. Lastly, In section \ref{Sect4} we discuss some possible configurations which develop singularities within the Einstein-Skyrme system and we give some outlooks for further study in the last section, namely, Section \ref{sec:concl}.
\section{The Einstein-Skyrme System in Bondi Coordinate}\label{Sect2}
We specifically consider a spherically symmetric spacetime for the model. The metric of the spacetime is written in Bondi coordinate that is given by \cite{Bondi:1962px}
\begin{equation}
    ds^2=\gamma_{\mu\nu}dx^\mu dx^\nu=-e^{2\rho}du^2-2e^{\rho+\sigma}dudr+r^2d\Omega^2,
\end{equation}
where \(\rho\equiv\rho(u,r)\) and \(\sigma\equiv\sigma(u,r)\), both are functions of \(u\) and \(r\) and \(d\Omega^2=d\theta^2+\sin^2\theta d\varphi^2\) is the metric of 2-sphere, \(S^2\). Einstein's summation convention is to be understood for every repeated indices throughout this paper unless stated otherwise. The coordinate \(r\) is the radial coordinate which is defined through the area of spatial rotation \(SO(3)\) group orbit, \(\mathcal{A}\), that satisfies \(\mathcal{A}=4\pi r^2\), implying that the central world-line is the set of points \(r=0\). The coordinate \(u\) is defined by requiring it to be constant on the future light cone of each point on the central world-line and \(u\) tends to the proper time on the world line of constant \(r\) as \(r\) goes to infinity \cite{Christodoulou:1986zr,Christodoulou:1986du}. In this way, we can consider the coordinate \(u\) as the  retarded time. The null tetrad in this coordinate consists of two orthonormal bases on \(S^2\) and two vectors
\begin{equation}
    n=e^{-\rho}\frac{\partial}{\partial u}-\frac{1}{2}e^{-\sigma}\frac{\partial}{\partial r},~~~l=e^{-\sigma}\frac{\partial}{\partial r},
\end{equation}
both \(n\) and \(r\) are orthogonal to the bases on \(S^2\) \cite{Christodoulou:1986zr,Christodoulou:1994hg}.

The Einstein equation of the Einstein-Skyrme system that models the Skyrme field minimally coupled to gravity is given by
\begin{equation}\label{Einstein}
    R_{\mu\nu}=8\pi\left[ T_{\mu\nu}-\frac{1}{2}\gamma_{\mu\nu}T\right]=8\pi\mathcal{T}_{\mu\nu},
\end{equation}
with \(R_{\mu\nu}\) is the Ricci tensor corresponding to the metric tensor \(\gamma_{\mu\nu}\), \(R=R_{\mu\nu}\gamma^{\mu\nu}\) is the Ricci scalar, and \(T_{\mu\nu}\) is the energy-momentum tensor of matter fields, which, in this case, is in the form of Skyrme field.

In order to find an explicit expression of the energy-momentum tensor of Skyrme fields, we consider the action of the Skyrme model, that is expressed in equation \eqref{SkyrmeAction}. This model contains a real scalar multiplet, \(\phi^a=(\phi^0,\phi^1,\phi^2,\phi^3)\) which satisfies the \(O(4)\) model constraint, \(\phi^a\phi^a=1\). This constraint implies that the Skyrme field maps the spacetime manifold to a three-dimensional sphere \(S^3\). Here, we have changed the representation of the Skyrme field to \(O(4)\) from the usual \(SU(2)\) valued field to simplify the calculation, without any loss of generality, by following the construction of \(O(N)\) Skyrme-Sigma Models (see \cite{Arthur:1996ia} for details).  Since the target space is \(S^3\), which is the same target space manifold of an \(SU(2)\) valued field, then the model obeys the same symmetry. In fact, the relationship between \(\phi\) and the usual \(SU(2)\) valued field is straight forward, namely
\begin{equation}
    U=\phi^0 \textbf{1}_{2\times2}+i(\phi^1\sigma^1+\phi^2\sigma^2+\phi^3\sigma^3),
\end{equation}
where \(\textbf{1}_{2\times2}\) is the \(2\times2\) identity matrix and \((\sigma^1,\sigma^2,\sigma^3)\)
are the Pauli matrices. We can see that the \(O(4)\) model constraints, \(\phi^a\phi^a=1\), is equivalent to \(U^\dagger U=\textbf{1}_{2\times2}\).

With the action of the Skyrme model in hand, we can derive the energy-momentum tensor corresponding to Skyrme field by employing the following equation
\begin{equation}
    T_{\mu\nu}=-\frac{2}{\sqrt{-\gamma}}\frac{\delta \mathcal{S}_{\text{Skyrme}}}{\delta \gamma^{\mu\nu}},
\end{equation}
where \(\delta\) is the variational operator. This formula will ensure that Einstein's equations are equivalent to \(\frac{1}{\sqrt{-\gamma}}\frac{\delta \mathcal{S}}{\delta \gamma^{\mu\nu}}=0\). As such, the expression of energy-momentum tensor for the Skyrme field is as follows
\begin{eqnarray}
    T_{\mu\nu}=2\left[C_1\phi^a_\mu\phi^a_\nu+2C_2\phi^a_{[\mu}\phi^b_{\alpha]}\phi^a_{[\nu}\phi^b_{\beta]}\gamma^{\alpha\beta}\right]-\gamma_{\mu\nu}\left[C_1\phi^a_\kappa\phi^a_\tau \gamma^{\kappa\tau}+C_2\phi^a_{[\kappa}\phi^b_{\alpha]}\phi^a_{[\tau}\phi^b_{\beta]}\gamma^{\kappa\tau}\gamma^{\alpha\beta}\right].
\end{eqnarray} 
 From equation \eqref{Einstein} we can see that \(T_{\mu\nu}\) must have the same dimensions as \(R_{\mu\nu}\). Because scalar multiplets \(\phi^a\) are dimensionless quantities (see, \cite{Skyrme:1961vq,manton1987,Brihaye:2017wqa}), then \(C_1\) must be a dimensionless constant and \(C_2\) must have a dimension of length squared. This analyses of dimension for coupling constants is useful for constructions of dimensionless equation of motion we show in later sections. The Einstein-Skyrme system is supplemented by matter field equation, given by
\begin{equation}\label{matterEq}
    \nabla_\mu T^{\mu\nu}=0,
\end{equation}
such that we have all dynamical equations for both the metric tensor and Skyrme (matter) field.

The original Skyrme model, firstly proposed by T. Skyrme in \cite{Skyrme:1961vq,Skyrme:1962vh}, does not accommodate potential. Such a model has minimal terms to be considered as stable in the sense that Derrick stability conditions are satisfied and the lack of potential term implies that the corresponding Skyrmion is massless \cite{Manton:2004tk}. It has been proven that this model satisfies BPS (Bogomol'nyi–Prasad–Sommerfield) inequality \(E\geq B \) where $B$ is a constant topological charge and it becomes equality if the map \(\phi\) is given by \(S^3\rightarrow S^3\), implying a spherical base manifold \cite{manton1987}. Thus, the original Skyrme model is topologically supported\footnote{In the sense that the energy cannot go to zero for higher topological degree, \(B\geq 1\), hence, skyrmions cannot disperse into vacuum spontaneously} even though it comes from different mechanism from the BPS Skyrme model and the generalized Skyrme model.

For the geometric features of the spacetime of our model, we start by showing the corresponding Ricci tensor. The components of the Ricci tensor of this Bondi coordinate are given by
\begin{eqnarray}
    R_{uu}&=&\frac{e^{\rho- \sigma }}{r}\left(e^{\rho-\sigma } \left(\rho_{,r}\left(r \rho_{,r}-r \sigma_{,r}+2\right)+r \rho_{,rr}\right)- \left(r \left(\rho_{,ur}+\sigma_{,ur}\right)+2 \sigma _{,u}\right)\right),\\
    R_{ur}&=&R_{ru}=\frac{e^{\rho-\sigma}}{r}\left(\rho_{,r} \left(r \rho_{,r}-r \sigma_{,r}+2\right)+r
   \rho_{,rr}\right)-\rho_{,ur}-\sigma _{,ur},\\
   R_{rr}&=&\frac{2}{r}\left(\rho_{,r}+\sigma_{,r}\right),\\
   R_{\theta\theta}&=&1-e^{-2 \sigma} \left(r \rho _{,r}-r \sigma _{,r}+1\right),\\
   R_{\varphi\varphi}&=&\sin^2\theta\left(1-e^{-2 \sigma} \left(r \rho _{,r}-r \sigma _{,r}+1\right)\right).
\end{eqnarray}

In this work, we use spherically symmetric ansatz possessing unit topological charge, known as the hedgehog ansatz, given by
\begin{eqnarray}\label{ansatzPhi}
    &&\phi^0=\cos\xi,~~~~~~~~~~~~~~~~~~~\phi^1=\sin\xi\cos(A_3\theta),\nonumber\\
    &&\phi^2=\sin\xi\sin(A_3\theta)\cos(A_4\varphi),~~\phi^3=\sin\xi\sin(A_3\theta)\sin(A_4\varphi),
\end{eqnarray}
with  \(\xi\equiv\xi(u,r)\) is a real valued function, $A_3, A_4 \in\mathbb{Z}$ to ensure the uniqueness of \(\phi\) in every point in spacetime, and both \(\theta\) and \(\varphi\) take values in the usual range for angular coordinates on \(S^2\), namely \(\theta\in[0,\pi)\) and \(\varphi\in [0,2\pi)\). The boundary conditions for \(\xi\) also come from the requirement that \(\phi\) must be unique, then at \(r=0\), \(\sin \xi\) must be zero which leads to \(\xi(u,0)=n\pi\) where \(n\in\mathbb{Z}\). We restrict the solutions, from this point on to the decaying solutions of \(\xi\) with respect to \(r\) such that we always assume \(\lim_{r\rightarrow\infty}\xi(u,r)=0\), and then we show the proof that such smooth decaying solutions exist in the following sections.
Now, we proceed by substituting this ansatz into \(\mathcal{T}_{\mu\nu}=T_{\mu\nu}-\frac{1}{2}\gamma_{\mu\nu}T\) to have the following results
\begin{eqnarray}
   \mathcal{T}_{rr}&=&\left[C_1\left(2\xi_{,r}^2\right)+C_2\left(2\left(A_3^2+A_4^2\frac{\sin^2(A_3\theta)}{\sin^2\theta}\right)\frac{\sin^2\xi}{r^2}\xi_{,r}^2\right)\right]~,\\
   \mathcal{T}_{\theta\theta}&=&C_2A_3^2A_4^2\frac{\sin^2(A_3\theta)}{\sin^2\theta}\frac{\sin^4\xi}{r^2}+C_1\left(2A_3^2\sin^2\xi\right)\nonumber\\
   &&-C_2\left(A_3^2-A_4^2\frac{\sin^2(A_3\theta)}{\sin^2\theta}\right)\xi_{,r} \sin ^2\xi  e^{-\rho -2 \sigma } \left(2 \xi_{,u} e^{\sigma}-\xi_{,r} e^{\rho}\right),\nonumber\\
   \mathcal{T}_{\varphi\varphi}&=&\left[C_2A_3^2A_4^2\frac{\sin^2(A_3\theta)}{\sin^2\theta}\frac{\sin^4\xi}{r^2}+C_1\left(2A_4^2\frac{\sin^2(A_3\theta)}{\sin^2\theta}\sin^2\xi\right)\right.\nonumber\\
   &&\left.-C_2\left(A_3^2-A_4^2\frac{\sin^2(A_3\theta)}{\sin^2\theta}\right)\xi_{,r} \sin ^2\xi  e^{-\rho -2 \sigma } \left(2 \xi_{,u} e^{\sigma}-\xi_{,r} e^{\rho}\right)\right]\sin^2\theta.\nonumber
\end{eqnarray}
The angular components are singular at \(\theta\rightarrow0\) and \(\theta\rightarrow\pi\), thus in order to have a well-behaved energy-momentum tensor and a spherically symmetric \(rr\) component, we need to set \(|A_3|=1\). This leads to a specific value for \(A_4\), since in order to make \(\theta\theta\) and \(\varphi\varphi\) components of Einstein's equation consistent requires \(A_3^2-A_4^2=0\), then \(|A_4|=1\). This way, spherical symmetry is restored because \(\mathcal{T}_{\varphi}^\varphi=\mathcal{T}_{\theta}^\theta\).

By exploiting the same formulation for \(O(d+1)\) models in \cite{Arthur:1996ia,Brihaye:2017wqa,Fadhilla:2020rig}, we have the following expression of topological charge, given by
\begin{eqnarray}
    B&=&-\frac{1}{2\pi^2}\int \sqrt{-g}A_3A_4\frac{\sin(A_3\theta)}{r^2\sin\theta}e^{-(\rho+\sigma)}\sin^2\xi \xi' ~d\varphi d\theta dr\nonumber\\&=&\frac{A_3A_4}{2\pi^2}\int_0^{n\pi}\int_0^{\pi}\int_0^{2\pi}\sin^2\xi\sin(A_3\theta)~d\varphi d\theta d\xi.
\end{eqnarray}
Evaluating the integral expression above and using the argument, \(|A_3|=|A_4|=1\), in the preceding paragraph gives
\begin{equation}
    B= \pm n.
\end{equation} 
The spherically symmetric minimal energy Skyrmion has \(n=1\). 
From this point on, we are going to consider only positive \(n\) and positive signed \(B\) in order to simplify the analysis. The negative signed \(B\) are called the anti-Skyrmions and they behave similarly to the Skyrmions.

To proceed, we choose to take the \({rr}\) and \({\theta\theta}\) components of  \eqref{Einstein}, namely
\begin{eqnarray}\label{Einstein1}
    \frac{1}{r}\left(\rho_{,r}+\sigma_{,r}\right)&=&8\pi\xi_{,r}^2\left[C_1+C_2\left(2\frac{\sin^2\xi}{r^2}\right)\right]~,\\
    \frac{e^{2\sigma}-1}{r^2}-\frac{1}{r}\left(\rho_{,r}-\sigma_{,r}\right)&=&8\pi e^{2\sigma}\left[C_1\left(2\frac{\sin^2\xi}{r^2}\right)+C_2\left(\frac{\sin^4\xi}{r^4}\right)\right],\label{Einstein2}
\end{eqnarray}
and the rest of the equations from \eqref{Einstein} should be consistent with \eqref{Einstein1} and \eqref{Einstein2}. The \({u}\) and \({r}\) components of matter field equation \eqref{matterEq} is given by
\begin{eqnarray}\label{Ori}
    0&=&\left[\frac{2}{r^2}\left(r e^{-2 \sigma} \left(\xi_{,r} \left(-r \rho_{,r}+r \sigma_{,r}-2\right)+2 \left(\xi_{,u}+r \xi_{,ur}\right) e^{\sigma-\rho}-r \xi_{,rr}\right)+\sin (2 \xi )\right)\right]\nonumber\\
    &&+\left[\frac{4 \sin\xi}{r^4}\left( 2 r^2 e^{-\rho-\sigma} \left(\xi_{,ur} \sin\xi+\xi_{,r}\xi_{,u} \cos\xi\right)+\sin ^2\xi \cos \xi\right.\right.\nonumber\\
    &&\left.\frac{\left.\right.\left.\right.}{\left.\right.}r^2\left.e^{-2 \sigma} \left(\sin\xi \left(\xi_{,r} \left(\rho_{,r}-\sigma_{,r}\right)+\xi_{,rr}\right)+\xi_{,r}^2 \cos\xi\right)\right)\right],
\end{eqnarray}
and the rest of the components of \eqref{matterEq} are trivial. In equation \eqref{Ori} we have rescaled the coordinates as \((u,r)\rightarrow \left(\sqrt{\frac{C_1}{C_2}}u,\sqrt{\frac{C_1}{C_2}}r\right)\) such that the resulting equation becomes dimensionless. From the set of field equations above, we can see that the topological charge does not explicitly appear in the differential equations. Thus, the estimates for upper bound and lower bounds in our analysis will not depend on the topological charge. Instead, the dependency on the topological charge appears in the analysis of suitable initial data where boundary conditions play some important roles, especially for the lower bounds of the norm of initial data and the size of the interval of local existence that is going to be demonstrated in details in the next section.

\section{Existence of Smooth Solutions}\label{Sect3}
In this section, we show that some classes of smooth solutions exist in this system by using contraction mapping theorem. To do so, we recast the system of equations \eqref{gOri}, \eqref{gTildeOri}, and \eqref{Ori} into a single integro-differential equation as demonstrated below. 

Firstly, it is useful to introduce averaged field notation for any function, \(f\), of \(u,r\) on \(r\), given by
\begin{equation}
    \Bar{f}=\frac{1}{r}\int_0^rf(u,s)ds.
\end{equation}
Following the prescription on \cite{Christodoulou:1986zr,Chae:2001ec}, we introduce a new field \(h= h(u,r)\equiv \frac{\partial}{\partial r}\left(r\xi\right)\) which allows us to transform \(\xi\) and its derivative as
\begin{equation}
    \xi=\Bar{h},~~~\xi_{,r}=\frac{h-\Bar{h}}{r}.
\end{equation}
By defining another functions of \(u,r\) for metric fields, namely
\begin{eqnarray}\label{defg}
    g &\equiv& e^{\rho+\sigma},\\\label{defTildeg}
    \Tilde{g} &\equiv& e^{\rho-\sigma},
\end{eqnarray}
and introducing a differential operator parallel to \(n\)
\begin{equation}
    D=\frac{\partial}{\partial u}-\frac{\Tilde{g}}{2}\frac{\partial}{\partial r}
\end{equation}
we rewrite \eqref{Einstein1} and \eqref{Einstein2} as follows
\begin{eqnarray}\label{gOri}
    g&=&\exp\left[-8\pi C_1\int^\infty_r\frac{(h-\Bar{h})^2}{s}\left(1+2\frac{\sin^2\Bar{h}}{s^2}\right)~ds\right],\\
    \label{gTildeOri}\Tilde{g}&=&\Bar{g}-\frac{8\pi C_1}{r}\int_0^rs^2g\left(2\frac{\sin^2\Bar{h}}{s^2}+\frac{\sin^4\Bar{h}}{s^4}\right)~ds.
\end{eqnarray}
which are analogous to the relations we have for Einstein-Klein-Gordon system \cite{Chae:2001ec}. Now, with these two relations for \(g\) and \(\Tilde{g}\) in hand, we can proceed by reducing the system into a single integro-differential equation for \(h\) on a curve \(\chi(u',r')\) as a solution of 
\begin{equation}\label{ChiCurve}
   \frac{dr}{du}=-\frac{1}{2}\Tilde{g} 
\end{equation}
possessing tangential vector field which is parallel to \(D\). Thus, the characteristic curve is located on the null region of the spacetime.

Notice that after these rescaling, \(u,r\) become dimensionless parameters.
Let us define a new coupling constant between matter field and gravity, given by
\begin{equation}
    \alpha=8\pi C_1.
\end{equation}
Thus, by substituting \eqref{gOri} and \eqref{gTildeOri} to \eqref{Ori} we arrive at the following integro-differential equation
\begin{eqnarray}\label{DynEqOri}
    0&=&Dh-\frac{(h-\Bar{h})(g-\Tilde{g})}{2r}+\frac{\alpha}{2}(h-\Bar{h}) gr\left(2\frac{\sin^2\Bar{h}}{r^2}+\frac{\sin^4\Bar{h}}{r^4}\right)+\frac{g}{r}\sin\Bar{h}\cos\Bar{h}\nonumber\\
    &&+2\frac{\sin^2\Bar{h}}{r^2}Dh+2\frac{\sin\Bar{h}}{r}\left(hD\left(\frac{\sin\Bar{h}}{r}\right)-D\left(\sin\Bar{h}\frac{\Bar{h}}{r}\right)-\sin\Bar{h}\frac{(h-\Bar{h})(g-\Tilde{g})}{2r}\right.\nonumber\\
    &&+\left.\frac{\alpha}{2}\sin\Bar{h}(h-\Bar{h})g\left(2\frac{\sin^2\Bar{h}}{r^2}+\frac{\sin^4\Bar{h}}{r^4}\right)\right)+\frac{g}{r^3}\sin^3\Bar{h}\cos\Bar{h}
\end{eqnarray}
with only one dimensionless free coupling constant, \(\alpha\) to be tuned. 

Now, consider the solution of \eqref{DynEqOri} along the curve \(\chi\), then we can reduce the partial integro-differential equation into an ordinary one, parameterized by \(u'\), namely
\begin{equation}
    \frac{dh}{du'}=Ph+Q,
\end{equation}
with
\begin{eqnarray}
    P&\equiv&\left[\frac{g-\Tilde{g}}{2r}-\frac{\alpha}{2} g r\left(2\frac{\sin^2\Bar{h}}{r^2}+\frac{\sin^4\Bar{h}}{r^4}\right)\right]_\chi-\frac{1}{2}\frac{d}{du'}\ln\left(1+2\frac{\sin^2\Bar{h}}{r^2}\right)_\chi\\
    Q&\equiv&-\left[\Bar{h}\frac{g-\Tilde{g}}{2r}-\frac{\alpha}{2}\Bar{h}gr\left(2\frac{\sin^2\Bar{h}}{r^2}+\frac{\sin^4\Bar{h}}{r^4}\right)\right]_\chi+\left(\frac{2\frac{\sin^2\Bar{h}}{r^2}}{1+2\frac{\sin^2\Bar{h}}{r^2}}\right)_{\chi}\left.\frac{d}{du'}\Bar{h}\right|_\chi\nonumber\\
    &&-\left.g\cos\Bar{h}\frac{\sin\Bar{h}}{r}\right|_\chi\left(\frac{1+\frac{\sin^2\Bar{h}}{r^2}}{1+2\frac{\sin^2\Bar{h}}{r^2}}\right)_\chi+\frac{\Bar{h}}{2}\frac{d}{du'}\ln\left(1+2\frac{\sin^2\Bar{h}}{r^2}\right)_\chi.
\end{eqnarray}
Thus we can define a map \(h\rightarrow\mathcal{F}(h)\) as
\begin{equation}\label{F}
    \mathcal{F}(h)=h_0\exp\left(\int_0^{u'}Pdu\right)+\left[\int_0^{u'}Q\exp\left(\int_u^{u'}Pd\Tilde{u}\right)du\right],
\end{equation}
where \(h_0\equiv h_0(r)=h(0,r)\) is the initial data of \(h\).  With equation \eqref{F} in hand, the proof of local and global existence of the original Einstein-Skyrme system solution can be provided by showing that \(\mathcal{F}\) is a contractive mapping. It is worth noting that \(P\) and \(Q\) satisfy the following Inequality
\begin{eqnarray}
    \left|P\right|&\leq& \left|\frac{g-\Tilde{g}}{2r}\right|+\left|\frac{\alpha}{2} g r\left(2\frac{\Bar{h}^2}{r^2}+\frac{\Bar{h}^4}{r^4}\right)\right|+\frac{d}{du'}\left|2\frac{\Bar{h}^2}{r^2}\right|,\\
    \left|Q\right|&\leq&\left|\Bar{h}\frac{g-\Tilde{g}}{2r}\right|+\left|\frac{\alpha}{2} \Bar{h} g r\left(2\frac{\Bar{h}^2}{r^2}+\frac{\Bar{h}^4}{r^4}\right)\right|+\left|g\frac{\Bar{h}^3}{2r}\right|+\frac{d}{du'}\left|2\frac{\Bar{h}^3}{r^2}\right|
\end{eqnarray}
because \(\sin\Bar{h}\leq\Bar{h}\) and \(\frac{\sin{\Bar{h}}}{r}\geq0\) for every point on \(\chi\). The dynamical equation of first derivative of \(h\) can be found by taking the derivative of \eqref{DynEqOri} with respect to \(r\) to give
\begin{equation}
    \frac{d}{du'}\left(\frac{\partial h}{\partial r}\right)=P\left(\frac{\partial h}{\partial r}\right)+U,
\end{equation}
where \(U\) satisfies,
\begin{eqnarray}
    \left|U\right|&\leq&\left(\left|2\alpha B^2\left(7g+r\frac{\partial g}{\partial r}\right)\left(\frac{\Bar{h}^2}{r^2}+\frac{\Bar{h}^4}{r^4}\right)\right|+\left|\frac{1}{2r}\frac{\partial g}{\partial r}\right|+\frac{d}{du'}\left|10B\frac{\Bar{h}^2}{r^3}\right|\right)\left(h+\Bar{h}\right)\nonumber\\&&+\left(\left|\frac{g-\Tilde{g}}{2r^2}\right|+\left|\frac{4B}{r^2}\Bar{h}^2\left(g+r\frac{\partial g}{\partial r}\right)\right|\right)\left|h-\Bar{h}\right|.
\end{eqnarray}
Thus, by similar approach with estimation of \(h\) using \(\mathcal{F}\) \cite{Chae:2001ec}, we define a map \(\mathcal{G}:\frac{\partial h}{\partial r}\rightarrow \mathcal{G}\left(\frac{\partial h}{\partial r}\right)\) as follows
\begin{equation}\label{G}
    \mathcal{G}\left(\frac{\partial h}{\partial r}\right)=\frac{\partial h_0}{\partial r}\exp\left(\int_0^{u'}P~du\right)+\left[\int_0^{u'}\exp\left(\int_u^{u'}P~d\Tilde{u}\right)U~du\right].
\end{equation}
The reason we need an explicit form of the first derivative with respect to \(r\) is because we seek a solution that is at least \(C^1\), i.e. smooth solution in \(r\).

 Lastly, we define \(\Delta \equiv \mathcal{F}(h_2)-\mathcal{F}(h_1)\) such that the integro-differential equation of \(\Delta\) can be derived from \eqref{DynEqOri} to give
\begin{equation}\label{dynEqDelTilde}
    \frac{d}{du'}\Delta=P_2\Delta+(P_2-P_1)\mathcal{F}(h_1)+\frac{\Tilde{g}_2-\Tilde{g}_1}{2}\mathcal{G}( h_1)+(Q_2-Q_1),
\end{equation}
where \(P_1\equiv P(h_1)\), \(P_2\equiv P(h_2)\), \(Q_1\equiv Q(h_1)\), and \(Q_2\equiv Q(h_2)\). We are going to use the maps \(\mathcal{F}\) and \(\mathcal{G}\) in equations \eqref{F} and \eqref{G}, respectively, to find the radius of the open-ball where both maps map the elements into other elements inside the same ball, and then followed by estimating differential equations for \(\Delta\), equation \eqref{dynEqDelTilde} to find the corresponding Lipschitz condition. Thus, by employing Banach's fixed-point theorem, we can show that smooth classical solutions exist and, further, provide some conditions where the existence of solutions might fail within this approach.

\subsection{Local Existence Of Classical Solutions}
In this subsection, we prove the existence of the unique smooth local classical solutions to the Einstein-Skyrme system in Bondi coordinates for some specific initial data. Following similar method in \cite{Christodoulou:1986zr}, we define a function space
\begin{align}\label{spaceXtilde}
\Tilde{X}=\{h(.,.)\in C^1([0,u_0]\times [0,\infty))~|~\|h\|_{\Tilde{X}}<\infty \},
\end{align} 
with its norm \(\|h\|_{\Tilde{X}}\) defined as
\begin{align}\label{normXtilde}
\|h\|_{\Tilde{X}}:=\sup_{u\in [0,u_0]} \sup_{r\geq 0} \left\{(1+r)^{2}|h(u,r)|+(1+r)^3 \left|\frac{\partial h}{\partial r}(u,r)\right|\right\},
\end{align}
and a larger function space \(\Tilde{Y}\), containing \(\Tilde{X}\subset\Tilde{Y}\), defined as
\begin{align}\label{spaceYtilde}
\Tilde{Y}=\{h(.,.)\in C^1([0,u_0]\times [0,\infty))~|~\|h\|_{\Tilde{Y}}<\infty \},
\end{align} 
with 
\begin{align}\label{normYtilde}
\|h\|_{\Tilde{Y}}:=\sup_{u\in [0,u_0]} \sup_{r\geq 0} \left\{(1+r)^{2}|h(u,r)|\right\}.
\end{align}
The theorem of local existence is stated as follows
\begin{theorem}\label{TheoremLocal}
For every smooth initial data \(h_0(r)\in C^1[0,\infty)\) such that \(h_0(r)=O(r^{-2})\) and \(\frac{\partial h_0(r)}{\partial r}=O(r^{-3})\) as \(r\) goes to infinity, there exist \(u_0>0\) where unique smooth classical solution of spherically symmetric Einstein-Skyrme system with integer topological charge \(B>0\), \(h(u,r)\), exist on the interval \([0,u_0]\), taking \(h_0(r)=h(0,r)\) as its initial data. This solution behaves like  \(h(u,r)=O(r^{-2})\) and \(\frac{\partial h(u,r)}{\partial r}=O(r^{-3})\) as \(r\) goes to infinity for every \(u\) in \([0,u_0]\).
\end{theorem}

\textbf{Proof:} Suppose that our solution, \(h(u,r)\) is defined in an interval \(\mathcal{U}= [0,\eta]\) that contains \([0,u_0]\) such that \(\eta\in \mathcal{U}\) but \(\eta\notin [0,u_0]\). Let \(\Tilde{x}=\|h\|_{\Tilde{X}}\) and \(\Tilde{d}=\|h_0\|_{\Tilde{X}}=\sup_{r\geq0}\left\{(1+r)^{2}|h(0,r)|+(1+r)^3 \left|\frac{\partial h}{\partial r}(0,r)\right|\right\}\). From \eqref{gOri}, \eqref{gTildeOri}, and the fact that \(\frac{\partial g}{\partial r}=\frac{g-\Bar{g}}{r}\), we have the following estimate
\begin{eqnarray}
    \left|\frac{g-\Tilde{g}}{2r}\right|&\leq& \frac{1}{2}\left|\frac{\partial g}{\partial r}\right|+\frac{2\alpha}{r^2}\left|\int_0^rs^2g\left(2\frac{\sin^2\Bar{h}}{s^2}+\frac{\sin^4\Bar{h}}{s^4}\right) ds\right|\nonumber\\
     &\leq& \alpha  \frac{r(\Tilde{x}^2+\Tilde{x}^4)}{(1+r)^3}\leq \alpha (\Tilde{x}^2+\Tilde{x}^4).
\end{eqnarray}
As such, integral of \(P\) from \(u\in[0,u_0]\) up to \(u'=\eta>u_0\) can be estimated as
\begin{eqnarray}
    \int_u^\eta P d\Tilde{u}&\leq&\int_0^\eta \left|P\right| d\Tilde{u}\nonumber\\&\leq& \int_0^\eta \left|\alpha  \frac{r(\Tilde{x}^2+\Tilde{x}^4)}{(1+r)^3}+\alpha  \frac{\Tilde{x}^2+\Tilde{x}^4}{r^2(1+r)^2}\right| du + 2 \frac{\Tilde{x}^2}{r^2(1+r)^2}\nonumber\\
    &\leq& 2\alpha  (\Tilde{x}^2+\Tilde{x}^4)\eta.
\end{eqnarray}
Next, we deduce the following estimate for \(\mathcal{F}\) and \(\mathcal{G}\) from the estimate of the integral of \(P\) as follow
\begin{eqnarray}\label{estFlocal}
    \left|\mathcal{F}(\eta,r')\right|&\leq& \exp\left(K_1'(\Tilde{x}^2+\Tilde{x}^4)\eta\right)\left[\frac{\Tilde{d}}{(1+r')^2}+\int_0^{\eta}Qdu\right]\\
    \left|\mathcal{G}(\eta,r')\right|&\leq&  \exp\left(K_1'(\Tilde{x}^2+\Tilde{x}^4)\eta\right)\left[\frac{\Tilde{d}}{(1+r')^3}+\int_0^{\eta}Udu\right],\label{estGlocal}
\end{eqnarray}
with \(K_1'\geq 2\alpha \). Since \(\Tilde{g}\) is always less than one, every point \((u,r)\) satisfies \(r-r'\geq \frac{u'-u}{2}\) where \((u',r')\in\chi\), then it is straightforward to show that both integrals of \(Q\) and \(U\) up to \(u'=\eta\) in \eqref{estFlocal} and \eqref{estGlocal} can be estimated as 
\begin{eqnarray}
    \left|\int_0^{\eta}Qdu\right|&\leq& 2\alpha \frac{(\Tilde{x}^2+\Tilde{x}^4)\eta}{(1+r')^2},\\
    \left|\int_0^{\eta}Udu\right|&\leq& 15\alpha^2 \frac{(\Tilde{x}^3+\Tilde{x}^5+\Tilde{x}^7+\Tilde{x}^9)\eta}{(1+r')^3}.
\end{eqnarray}
As a consequence, the norm of \(\mathcal{F}\) in \(\tilde{X}\) satisfies the following inequality
\begin{equation}
    \|\mathcal{F}\|_{\Tilde{X}}\leq K_2'\left(2d+\Tilde{x}^2+\Tilde{x}^3+\Tilde{x}^4+\Tilde{x}^5+\Tilde{x}^7+\Tilde{x}^9\right)\eta\exp\left(K_1'(\Tilde{x}^2+\Tilde{x}^4)\eta\right),
\end{equation}
with \(K_2'\geq 15\alpha^2\).
Now, let us take an arbitrary \(\Tilde{x}_1>d\) and take \(\eta=\eta(\Tilde{x}_1,d)\) to be the solution of
\begin{equation}\label{etaSol}
    K_2'\left(2d+\Tilde{x}_1^2+\Tilde{x}_1^3+\Tilde{x}_1^4+\Tilde{x}_1^5+\Tilde{x}_1^7+\Tilde{x}_1^9\right)\eta\exp\left(K_1'(\Tilde{x}_1^2+\Tilde{x}_1^4)\eta\right)-\Tilde{x}_1=0.
\end{equation}
By choosing such value\footnote{An explicit solution of equation \eqref{etaSol} can be written in principal Lambert \(W\)-function, \(\mathcal{W}_0(x)\), as \(\eta=\frac{1}{K'_1(\Tilde{x}_1^2+\Tilde{x}_1^4)}\mathcal{W}_0\left(\frac{K'_1(\Tilde{x}_1^3+\Tilde{x}_1^5)}{K_2'\left(2d+\Tilde{x}_1^2+\Tilde{x}_1^3+\Tilde{x}_1^4+\Tilde{x}_1^5+\Tilde{x}_1^7+\Tilde{x}_1^9\right)}\right)\), see \cite{COR96} for details.} for \(\eta\), we ensure that \(\|\mathcal{F}\|_{\Tilde{X}} \leq \Tilde{x}_1\). In fact, the choice of \(\eta\) is not unique since it completely depends on how we estimate the norm of \(\mathcal{F}\) in \(\Tilde{X}\). Thus, there might be some other possible choice for \(\eta\). The conclusion of the above argument is summarized as follows
\begin{lemma}\label{lemmaLoc1}
For every \(\tilde{x}_1>d\), there exist \(\eta(\tilde{x_1},d)>0\) such that if \(u_0<\eta\) the map \(\mathcal{F}\) defined in \eqref{F} is contained in a closed ball of radius \(\tilde{x}_1\). The mentioned closed ball is a subspace of \(\tilde{X}\).
\end{lemma}

Now, let us assume that there are two different solutions, \(h_1,h_2\in \tilde{X}\), for \eqref{DynEqOri}. Suppose that these solutions are defined in the same interval \(\mathcal{U}= [0,\nu]\) that contains \([0,u_0]\) such that \(\nu\in \mathcal{U}\) but \(\nu\notin [0,u_0]\). By assuming that \(h_1\) and \(h_2\) have exactly the same initial data, the solution of \eqref{dynEqDelTilde} satisfies the following inequality
\begin{equation}\label{DelIneq}
    \Delta\leq\int_0^{\nu} \exp\left(\int_u^{\nu} P_2 d\Tilde{u}\right)\left[(P_2-P_1)\mathcal{F}(h_1)+\frac{\Tilde{g}_2-\Tilde{g}_1}{2}\mathcal{G}(h_1)+(Q_2-Q_1)\right]du
\end{equation}
\begin{equation}
    \leq\exp\left(\int_0^{\nu} P_2 d\Tilde{u}\right)\int_0^{\nu} \left|(P_2-P_1)\mathcal{F}(h_1)+\frac{\Tilde{g}_2-\Tilde{g}_1}{2}\mathcal{G}(h_1)+(Q_2-Q_1)\right|du\nonumber
\end{equation}
Let us define \(\tilde{y}\equiv\|h_2-h_1\|_{\tilde{Y}}\) and take \(\tilde{x}\) to be greater than \(\max \{\|h_1\|_{\tilde{
X}},\|h_2\|_X\}\). The exponential factor of the integrand on the right hand side of \eqref{DelIneq} satisfies a similar estimate with the one we found on Lemma \ref{lemmaLoc1}. Thus, we only need to find the upper-bound of the remaining three terms. The estimate of these three terms are given as follow
\begin{eqnarray}
    \int_0^\nu\left|Q_2-Q_1\right|du&\leq&K_3' \frac{(\tilde{x}^3+\tilde{x})y}{(1+r')^2}\nu,\\
    \int_0^\nu\left|(P_2-P_1)\mathcal{F}(h_1)\right|du&\leq&K_3'\exp\left(K_1'(\Tilde{x}^2+\Tilde{x}^4)\nu\right) \frac{(\tilde{x}^3+\tilde{x})(d+\Tilde{x}^2+\Tilde{x}^4)y}{(1+r')^2}\nu^2,\\
    \int_0^\nu\left|\frac{\Tilde{g}_2-\Tilde{g}_1}{2}\mathcal{G}(h_1)\right|du&\leq&K_3' \exp\left(K_1'(\Tilde{x}^2+\Tilde{x}^4)\nu\right) \frac{(\tilde{x}^4+\tilde{x}^2)(d+\Tilde{x}^3+\Tilde{x}^5+\Tilde{x}^7+\Tilde{x}^9)y}{(1+r')^3}\nu^2,\nonumber\\
\end{eqnarray} with \(K_3'\leq 12\alpha^3\). From here, we are able to deduce the Lipschitz condition for \(\Delta\) in \(\tilde{Y}\), namely
\begin{equation}
    \|\Delta\|_{\tilde{Y}}\leq \tilde{\mathcal{L}}(\tilde{x})y,
\end{equation}
where we have defined
\begin{eqnarray}\label{lipsLoc}
    \tilde{\mathcal{L}}(\tilde{x})&\equiv& K_3'\left[(\tilde{x}^3+\tilde{x})(d+\Tilde{x}^2+\Tilde{x}^4)+(\tilde{x}^4+\tilde{x}^2)(d+\Tilde{x}^3+\Tilde{x}^5+\Tilde{x}^7+\Tilde{x}^9)+(\tilde{x}^3+\tilde{x})\right]\nu^2\nonumber\\&&
    \exp\left(K_1'(\Tilde{x}^2+\Tilde{x}^4)\nu\right).
\end{eqnarray}
The function \eqref{lipsLoc} is a monotonically increasing function and is zero for \(\tilde{x}=0\).
We can always choose arbitrary value of \(\tilde{x}_1>d\) and take \(\nu\) as a function of \(\tilde{x}_1\) to be small enough such that \(0\leq \tilde{\mathcal{L}}(\tilde{x})<1\) for every \(\tilde{x}\in[0,\tilde{x}_1]\) to ensure that \(\mathcal{F}\) is a contraction mapping.
Thus, from the above Lipschitz condition we can conclude that
\begin{lemma}\label{lemmaLoc2}
For every \(\Tilde{x}_1 > d \) there exist \(\nu(x_1)>0\), such that if \(u_0<\nu(x_1)\) the map \(\mathcal{F}\) defined in \eqref{F} contracts in \(\tilde{Y}\).
\end{lemma}
From Lemma \ref{lemmaLoc1} and Lemma \ref{lemmaLoc2} we deduce that unique smooth solutions of \eqref{DynEqOri}, that decays faster than \(O(r^{-2})\) as \(r\rightarrow\infty\), exist in the interval \(u\in[0,u_0]\) where \(u_0<\min[\eta,\nu]\). One can observe that the existence interval, \([0,u_0]\), always exist for every \(d\) since we can always find either \(\eta\) or \(\nu\) such that \(\Tilde{x}_1>d\) and then proceed to find the corresponding \(u_0\). 

We would also like to remark that there exists a lower bound of the norm of initial data in \(\Tilde{X}\).
\begin{lemma}
    The norm of initial data, \(d\equiv\|h_0\|_{\Tilde{X}}\) is bounded from below by,
    \begin{equation}
        d\geq n\pi,
    \end{equation}
    where \(n\in\mathbb{Z}\) is related to the topological charge, \(B=\pm n\).
\end{lemma}
\textbf{Proof: }The proof for the lower bound shown above is straightforward. Since \(\xi(u,0)=n\pi\), then
\begin{eqnarray}
    d&=&\sup_{u\in [0,u_0]} \sup_{r\geq 0} \left\{(1+r)^{2}|h_0(r)|+(1+r)^3 \left|\frac{\partial h_0(r)}{\partial r}\right|\right\}\nonumber\\
    &\geq&\lim_{r\rightarrow0}\left\{(1+r)^{2}|h_0(r)|+(1+r)^3 \left|\frac{\partial h_0(r)}{\partial r}\right|\right\}\geq h_0(0)\nonumber\\
    &=&\xi(0,0)=n\pi.
\end{eqnarray}
This feature implies that \(\tilde{x}_1>n\pi\) and implies that both \(\eta\) and \(\nu\) are finite because from both equations (\ref{etaSol}, \ref{lipsLoc}) and the fact that \(\tilde{\mathcal{L}}(\tilde{x})<1\) for contraction mapping, we have 
\begin{remark}
    \begin{eqnarray}
    \eta&<&\frac{1}{4\alpha n^4\pi^4}\mathcal{W}_0\left(\frac{1}{60\alpha n^4\pi^4}\right),\\
    \nu&<&\frac{1}{\alpha n\pi}\mathcal{W}_0\left(2\sqrt{3}\alpha^{5/2}\right),
\end{eqnarray}
where \(\mathcal{W}_0(x)\) is the principal Lambert's \(W\)-Function.
\end{remark}
 As such, we cannot extend these results to \(u_0\rightarrow\infty\). This means that we need a different approach in order to find global smooth solutions and some conditions for restricting the initial data are expected.

\subsection{Global Existence and Uniqueness Of Classical Solutions}
Here, we show that a globally unique solution exists for an Einstein-Skyrme system in Bondi coordinate with restricted initial data. To do so, by following the methods in\cite{Chae:2001ec, Christodoulou:1986zr} we introduce the function space \(X\),
\begin{align}\label{spaceX}
X=\{h(.,.)\in C^1([0,\infty)\times [0,\infty))~|~\|h\|_X<\infty \},
\end{align} 
with \(\|h\|_X\) defined as
\begin{align}\label{norm x}
\|h\|_X :=\sup_{u\geq 0} \sup_{r\geq 0} \left\{(1+r+u)^{2}|h(u,r)|+(1+r+u)^3 \left|\frac{\partial h}{\partial r}(u,r)\right|\right\}.
\end{align}
 We also define the function space 
\begin{align}\label{spaceX0}
X_0 = \{h(.)\in C^1([0,\infty))~|~\|h\|_{X_0}<\infty \},
\end{align}
with the norm
\begin{align}\label{norm x0}
\|h\|_{X_0}:=\sup_{r\geq 0}\left\{(1+r)^{2}|h(r)|+(1+r)^3\left|\frac{\partial h}{\partial r}(r) \right| \right\}
\end{align}
and we define the notation $d=\|h_0\|_{X_0}$. Furthermore, we define the function space \(Y\) containing \(X\),
\begin{align}\label{spaceY}
Y=\{h(.,.)\in C^1([0,\infty))\times[0,\infty))~|~h(0,r)=h_0(r),~\|h\|_Y<\infty\},
\end{align}
with the norm
\begin{align}\label{norm y}
\|h\|_Y = \sup_{u\geq 0}\sup_{r\geq 0}\left\{(1+r+u)^{2}|h(u,r)| \right\}.
\end{align}
As such, the statement of the theorem is as follows:
\begin{theorem}\label{TheoremGlobal}
Suppose we have initial data \(h_0\in X_0\) where \(X_0\) defined in \eqref{spaceX0}. Let \(d\) be the norm of \(h_0\), \(\|h_0\|_{X_0}\), and let \(X\) be a space defined in \eqref{spaceX}. There exists positive \(\alpha\) and \(\delta>n\pi\) such that if \(n\pi\leq d<\delta\), there exists a unique global classical solution \(h(u,r)\in X\) of Einstein-Skyrme system with integer topological charge \(B>0\), possessing a nonlinear dynamical equation \eqref{DynEqOri} with \(h_0(r)\) as its initial data. The decay properties of the solution are
\begin{eqnarray}
    \left|h(u,r)\right|&\leq& C(1+r+u)^{-2},\\
    \left|\frac{\partial h}{\partial r}(u,r)\right|&\leq& C(1+r+u)^{-3},
\end{eqnarray}
and the topological charge, \(B\), satisfies \(B=\pm n\), where \(n\in\mathbb{Z}\). The coupling constant is bounded from above by \(\alpha<\frac{58\mathcal{W}_0(\frac{3}{29})}{n^2\pi^2+n^4\pi^4}\).
\end{theorem}
In order to prove the theorem given above, first we need to define the map from \(X\) to the solution of \eqref{DynEqOri}, \(\mathcal{F}\), such that \(\mathcal{F}(h)\) solves \eqref{DynEqOri}. Thus, we need these two properties to be shown, namely
\begin{enumerate}
    \item \(\mathcal{F}:U\rightarrow U\), with \(U\subset X\).
    \item for \(h_1,h_2\in U \subset Y\) there exist \(\lambda\in(0,1)\) such that 
    \begin{eqnarray}
        \|\mathcal{F}(h_2)-\mathcal{F}(h_1)\|_Y\leq\lambda \|h_2-h_1\|_Y,
    \end{eqnarray}
    i.e. \(\mathcal{F}\) contracts in \(Y\).
\end{enumerate}
As mentioned earlier, both of the conditions above lead to the existence of unique solutions by using Banach's fixed-point theorem. Then, we are going to use such conditions to find suitable initial data for global solutions to exist. Furthermore, we can also study the influence of topology on the restrictions for initial data from the conditions above.

Firstly, we exploit the following estimates: 
\begin{itemize}
    \item \begin{equation}
    \left|\Bar{h}\right|\leq \frac{x}{(1+u)(1+u+r)}
\end{equation}
where \(x \equiv \|h\|_X\), and
\item \begin{equation} 
    \left|h(u,r)-h(u,r')\right|\leq \frac{x}{2}\left(\frac{1}{(1+u+r')^2}-\frac{1}{(1+u+r)^2}\right).
\end{equation}
\item As a consequence of both estimates above, we have
\begin{equation}
    \left|h-\Bar{h}\right|\leq \frac{xr}{2(1+r+u)^2(1+u)}
\end{equation}
\end{itemize}
 These estimates are needed to find the estimate of \(g\) and \(\Tilde{g}\). Similar to the space \(\Tilde{X}\) we use for the study of local solutions, we also have the lower bound for the norms of space \(X\) and \(X_0\).
 \begin{lemma}\label{GlobalLowBound}
      Let \(x=\|h(r)\|_X\) and \(d=\|h_0\|\), where \(h_0\) is the initial data of \(h\). Both \(x\) and \(d\) are bounded from below,
      \begin{equation}
          x\geq n\pi,~~~d\geq n\pi.
      \end{equation}
      \(n\in\mathbb{Z}\) is related to the topological charge by \(B=\pm n\).
 \end{lemma}
 \textbf{Proof: } Since the boundary condition of topological Skyrmions at the origin of space is \(\xi(u,0)=n\pi\), then \(h(u,0)=n\pi\) and \(h_0(0)=n\pi\). Thus, both \(h\) and \(h_0\) satisfies
 \begin{eqnarray}
     \|h\|_X&\geq& \lim_{r\rightarrow 0}\left((1+r)^{2}|h(u,r)|+(1+r)^3 \left|\frac{\partial h}{\partial r}(u,r)\right|\right)\geq h(u,0)=n\pi,\\
     \|h_0\|_{X_0}&\geq& \lim_{r\rightarrow 0}\left((1+r)^{2}|h_0(r)|+(1+r)^3 \left|\frac{\partial h_0}{\partial r}(r)\right|\right)\geq h_0(0)=n\pi.
 \end{eqnarray}
 This is the end of the proof of Lemma \eqref{GlobalLowBound}.

We should expect that the lower bounds for both norm of \(h\) and the norm of \(h_0\) above can be used to find the restriction for the initial data of global solutions. From previous subsections, we know that in order to have local solutions with arbitrary initial data, then the interval for the time coordinate \(u\) cannot be extended to infinity and bounded from above due to the topology of the Skyrmion. As such, the strategy to have global solutions, where the interval of \(u\) is extended to infinity, is that we need to restrict the norm of initial data, \(d\), and we expect that the value of \(d\) falls in an interval where the lower bound is given in Lemma \ref{GlobalLowBound}. Thus, we need to find the upper bound of \(d\) from the conditions of existence and uniqueness.

Now, consider the estimates for metric functions, \({g,\Tilde{g}}\), below.
\begin{proposition} \label{Prop1}
If \(h(u,r)\) belong to the space \(X\) defined in \eqref{spaceX} with initial data \(h(0,r)\) belong to the space \(X_0\) defined in \eqref{spaceX0}, then for two functions \({g,\Tilde{g}}\), given in \eqref{gOri} and \eqref{gTildeOri}, the following inequalities hold
\begin{eqnarray}
    g(u,r)&\geq&\exp\left[-\alpha \frac{\|h\|_X^2+\|h\|_X^4}{58}\right],\\
    \Tilde{g}(u,r)&\geq&\exp\left[-\alpha \frac{\|h\|_X^2+\|h\|_X^4}{58}\right]-\frac{\alpha }{6}\left(\|h\|_X^2+\|h\|_X^4\right).
\end{eqnarray}
Furthermore, we have
\begin{equation}
    |g-\Tilde{g}|\leq \alpha  \frac{(\|h\|_X^2+\|h\|_X^4)r^2}{6(1+u)^3(1+r+u)^3}
\end{equation}
\end{proposition}

\textbf{Proof:} At \(r\rightarrow0\) we have to impose the boundary condition \(\xi(r\rightarrow0)=n\pi\). Thus, we can estimate \(\sin\Bar{h}\) as \(\sin\xi=\sin(\pi-\xi)\leq\pi-\xi\leq -r\xi_{,r}=-(h-\Bar{h})\).
Consider the following estimate
\begin{eqnarray}
    \int^\infty_0\frac{(h-\Bar{h})^2}{s}\left(1+2\frac{\sin^2\Bar{h}}{s^2}\right)~ds&\leq& \int^\infty_0\frac{(h-\Bar{h})^2}{s}ds+2 \int^\infty_0\frac{(h-\Bar{h})^4}{s^3}ds\nonumber\\
    &\leq& \frac{x^2+x^4}{24},
\end{eqnarray}
which can be used to estimate the function \(g\) in \eqref{gOri} to give
\begin{equation}
    g\geq \exp\left[-\alpha \frac{x^2+x^4}{58}\right].
\end{equation}
Now, for \(\Tilde{g}\) we need the following estimate
\begin{eqnarray}
    \int_0^rs^2g\left(2\frac{\sin^2\Bar{h}}{s^2}+\frac{\sin^4\Bar{h}}{s^4}\right)~ds&\leq& 2\int_0^rs^2\left(\frac{(h-\Bar{h})^2}{s^2}+\frac{(h-\Bar{h})^4}{s^4}\right)ds\nonumber\\
    &\leq&\frac{(x^2+x^4)r^3}{6(1+u)^3(1+r+u)^3}.
\end{eqnarray}
By using equation \eqref{gTildeOri} with the estimate given above, we have
\begin{eqnarray}
    \Tilde{g}&\geq& \exp\left[-\alpha \frac{x^2+x^4}{58}\right]-\frac{\alpha}{r}\frac{(x^2+x^4)r^3}{6(1+u)^3(1+r+u)^3}\nonumber\\
    &\geq&\exp\left[-\alpha \frac{x^2+x^4}{58}\right]-\frac{\alpha }{6}\left(x^2+x^4\right).
\end{eqnarray}
Next, we estimate the following quantity
\begin{eqnarray}
    \int_{r'}^r \left|\frac{\partial g}{\partial s}(u,s)\right|ds &\leq& \frac{\alpha}{2} \int_{r'}^r \frac{(h-\Bar{h})^2}{s}\left(1+2\frac{(h-\Bar{h})^2}{s^2}\right)ds\nonumber\\
    &\leq&\frac{\alpha }{24 (1+u)^2}\left[x^2\left(\frac{1+3r'+u}{(1+r'+u)^3}-\frac{1+3r+u}{(1+r+u)^3}\right)\right.\nonumber\\&&+\left.x^4\left(\frac{1+7r'+u}{(1+r'+u)^7}-\frac{1+7r+u}{(1+r+u)^7}\right)\right]\nonumber\\
\end{eqnarray}
which is useful for estimate of \(\left|g-\Bar{g}\right|\) as follows
\begin{eqnarray}
    \left|g-\Bar{g}\right|&\leq& \frac{1}{r}\int_0^r\int_{r'}^r\left|\frac{\partial g}{\partial s}(u,s)\right|ds dr'\nonumber\\
   &\leq& \alpha \frac{(x^2+x^4)r^2}{12(1+u)^3(1+r+u)^3}.
\end{eqnarray}
From equation \eqref{gTildeOri}
we know that \(g-\Tilde{g}=g-\Bar{g}+\frac{\alpha}{r}\int_0^rs^2g\left(2\frac{\sin^2\Bar{h}}{s^2}+\frac{\sin^4\Bar{h}}{s^4}\right)~ds\). Thus, the following inequalities hold
\begin{eqnarray}
    \left|g-\Tilde{g}\right|&\leq& \left|g-\Bar{g}\right|+\frac{\alpha}{r}\left|\int_0^rs^2g\left(2\frac{\sin^2\Bar{h}}{s^2}+\frac{\sin^4\Bar{h}}{s^4}\right)~ds\right|\nonumber\\
    &\leq& \alpha  \frac{(x^2+x^4)r^2}{6(1+u)^3(1+r+u)^3}.
\end{eqnarray}
This completes the proof of Proposition \ref{Prop1}.

There are some implications from proposition \ref{Prop1}. Firstly, a black hole solution, where the lower bound of \(\Tilde{g}\) is less than zero,  leads to an additional condition for \(\|h\|_X\) and \(\|h\|_X\) is bounded from below. However, we can always find black hole solutions for every value of \(B\) since the lower bound of \(\Tilde{g}\) also depends on \(\alpha\), hence, the choice of topological degree does not affect the existence of black hole solutions but do affect the range of possible values for the effective gravitational coupling constant, \(\alpha\). Secondly, we can directly see that this bound is connected to the flat spacetime case, \(g=\Tilde{g}=1\) by choosing the matter field to be vacuum \(\|h\|_X=0\) but this can only be done for non-topological Skyrmions (\(B=0\)) or non-gravitating case (\(\alpha=0\)). Lastly, we would also like to remark that both \(g\) and \(\Tilde{g}\) converge into a single value at spatial infinity and time infinity which implies that the well-known static configurations (static Skyrme stars and Skyrme Blackhole) are the steady-state limit of this dynamical model. 

From here, we define a new constant 
\begin{equation}\label{k}
    k=\exp\left[-\alpha \frac{x^2+x^4}{58}\right]-\frac{\alpha }{6}\left(x^2+x^4\right),
\end{equation}
as a function of \(x\) defined for \(x\in[n\pi,x_0)\) where \(x_0\) is the root of \(k(x)=0\), and \(k(x)\) taking values from \((0,1]\) which is deduced from the possible values for the lower bound of \(\Tilde{g}\). The interval of \(k\) from the physical argument for \(\Tilde{g}\) leads to a restriction for the coupling constant \(\alpha\). Since we know that \(x\geq n\pi\), then in order to guarantee \(k\) falls in the interval \((0,1]\), we need 
\begin{equation}\label{intAlpha}
    0\leq\alpha<\frac{58\mathcal{W}_0(\frac{3}{29})}{n^2\pi^2+n^4\pi^4}.
\end{equation}
One shall notice that the higher the topological charge means that the interval for \(\alpha\) becomes smaller. Furthermore, we shall use the constant \(k\) to simplify our estimates.
The characteristic \eqref{ChiCurve} gives the following relation 
\begin{equation}
    r=r'+\frac{1}{2}\int_u^{u'}\Tilde{g}(\Tilde{u},r)d\Tilde{u}\geq r'+\frac{k}{2}(u'-u).
\end{equation}
From the inequality given above, we can deduce that \(1+r+u\geq \frac{k}{2}(1+r'+u')\) because \(1\geq k> 0\). There is also a useful relation for this characteristic which can be found in \cite{Chae:2001ec}, namely
\begin{equation}
    \int_0^{u'}\frac{r^s}{(1+u)^t(1+r+u)^q}du\leq \frac{2^m}{(t+q-s-m-1)k^m(1+r'+u')^m}
\end{equation}
for \(t+q-s-m>1\), where \(t,q,s,m\) are positive integers.

Now that we have the inequalities from characteristic, we can estimate \eqref{F} on the curve \(\chi\). Firstly, consider the first term in \eqref{F}. It is a product of initial condition \(h_0\) and the integrating factor. Because we can estimate \(h_0\) as 
\begin{equation}
    h_0\leq \frac{d}{(1+r(0))^2} \leq \frac{4d}{k^2(1+r'+u')^2},
\end{equation}
where we have used the fact that \(r(0)\geq r'+\frac{k}{2}u'\) and \(d\equiv \|h_0\|_{X_0}\), then the integrating factor can be estimated by exploiting the following relation
\begin{eqnarray}
    \int_0^{u'}Pdu &\leq& \int_0^{u'}\left|\frac{g-\Tilde{g}}{2r}\right|du+\int_0^{u'}
    \left|\frac{\alpha}{3} g r\left(2\frac{\Bar{h}^2}{r^2}+B^2\frac{\Bar{h}^4}{r^4}\right)\right|du+\left|2\Bar{h}^2\right|\nonumber\\
    &\leq&K_1 (x^2+x^4),
\end{eqnarray}
where satisfies \(K_1\geq 4\alpha\), which leads us to \(\exp\left(\int_0^{u'}Pdu\right)\leq \exp(K_1 (x^2+x^4))\). Now, for the second term, consider the following estimates
\begin{eqnarray}
    \int_0^{u'}Q\exp\left(\int_u^{u'}Pd\Tilde{u}\right)du &\leq& \exp(K_1 (x^2+x^4)) \left[\int_0^{u'}\left(\left|\Bar{h}\frac{g-\Tilde{g}}{2r}\right|+\left|\frac{\alpha}{2} \Bar{h} g r\left(2\frac{\Bar{h}^2}{r^2}+\frac{\Bar{h}^4}{r^4}\right)\right|\right.\right.\nonumber\\
    &&+\left.\left|g\frac{h-\Bar{h}}{r}\right|\right)du+\left.2\frac{\Bar{h}^3}{r^2}\right]\nonumber\\
    &\leq&  K_1\frac{x^3+x^5}{k^2(1+r'+u')^2}\exp(K_1 (x^2+x^4)).
\end{eqnarray}
Hence, we can conclude that \(\mathcal{F}(u',r')\) defined in \eqref{F} satisfies the following decay estimate
\begin{equation}\label{DecayF}
    \left|\mathcal{F}(u',r')\right|\leq K_1\frac{d+x^3+x^5}{k^2(1+r'+u')^2}\exp(K_1 (x^2+x^4)),
\end{equation}
i.e. \(\mathcal{F}\) is a bounded map whose upper bound is a decaying function that goes like \((1+r'+u')^{-2}\). All smooth functions, which decay sufficiently slower than the upper bound at \(r'\rightarrow\infty\), falls into the family of solutions for this case.

Next, we need to estimate the first derivative of \(h\) as well. 
To estimate \(\mathcal{G}\) in \eqref{G}, firstly we need to estimate the first derivative of initial condition \(h_0\), thus we have
\begin{eqnarray}
    \frac{\partial h_0}{\partial r}\leq \frac{d}{(1+r(0))^3}\leq \frac{8d}{k^3(1+r'+u')^3}.
\end{eqnarray}
Since the first derivative of \(h_0\) goes like \(O((1+r)^{-3})\) then we should adopt the same estimate of \(\exp\left(\int_0^{u'}Pdu\right)\) as the previous case for \(\mathcal{F}\). It is also necessary to find the estimate of the first derivative of \(g\) which is given by
\begin{eqnarray}
    \frac{\partial g}{\partial r}&=&g\alpha\frac{(h-\Bar{h})^2}{2r}\left(1+2\frac{\sin^2\Bar{h}}{r^2}\right)\nonumber\\
    &\leq& \alpha  \frac{(x^2+x^4)r}{4(1+r+u)^4(1+u)^2}.
\end{eqnarray}
With the estimate of \(\frac{\partial g}{\partial r}\) in hand we can proceed to find the estimate for \(\int_0^{u'}Udu\) as follows
\begin{eqnarray}
    \int_0^{u'}Udu&\leq&\int_0^{u'}\left(\left|\alpha \left(7g+r\frac{\partial g}{\partial r}\right)\left(\frac{\Bar{h}^2}{r^2}+\frac{\Bar{h}^4}{r^4}\right)\right|+\left|\frac{1}{2r}\frac{\partial g}{\partial r}\right|\right)\left(h+\Bar{h}\right)du\nonumber\\&&+\int_0^{u'}\left(\left|\frac{g-\Tilde{g}}{2r^2}\right|+\left|\frac{4}{r^2}\Bar{h}^2\left(g+r\frac{\partial g}{\partial r}\right)\right|\right)\left|h-\Bar{h}\right|du+10\left|\frac{\Bar{h}^4}{r'{}^3}\right|\nonumber\\
    &\leq& 14\alpha^2  \frac{x^3+x^4+x^5+x^7+x^9}{k^3(1+r'+u')^3}.
\end{eqnarray}
As such, the second term of \eqref{G} satisfies the inequality below
\begin{equation}
    \int_0^{u'}\exp\left(\int_u^{u'}Pd\Tilde{u}\right)Udu\leq K_2\frac{x^3+x^4+x^5+x^7+x^9}{k^3(1+r'+u')^3},
\end{equation}
where, \(K_2 \geq 14\alpha^2 \). As a conclusion, \(\mathcal{G}(u',r')\) defined in \eqref{G} is bounded from above by
\begin{equation}\label{DecayG}
    \left|\mathcal{G}(u',r')\right|\leq K_2\frac{d+x^3+x^4+x^5+x^7+x^9}{k^3(1+r'+u')^3}\exp\left(K_1(x^2+x^4)\right).
\end{equation}
We conclude the estimates for \(\mathcal{F}(u',r')\) and \(\mathcal{G}(u',r')\) in the following lemma
\begin{lemma}{(\textbf{Decay Properties of Einstein-Skyrme system})}\label{Prop2}
Suppose that \(h(u,r)\in X\) is the solution of \eqref{DynEqOri} with initial data \(h_0\in X_0\) and both \(\|h\|_X\) and \(\|h_0\|_{X_0}\) are finite, \(X\) and \(X_0\) are defined in \eqref{spaceX} and \eqref{spaceX0}, respectively. Then, \(h(u,r)\) and \(\frac{\partial h}{\partial r}(u,r)\) satisfy the following decay estimate:
\begin{eqnarray}
    \left|h(u,r)\right|&\leq& C(1+r+u)^{-2},\\
    \left|\frac{\partial h}{\partial r}(u,r)\right|&\leq& C(1+r+u)^{-3}.
\end{eqnarray}
\end{lemma}
\textbf{Proof:} If \(x=\|h\|_X\) and \(d=\|h_0\|_{X_0}\) are finite, then inequalities \eqref{DecayF} and \eqref{DecayG} states the upper bound for \(h(u,r)\) and \(\frac{\partial h}{\partial r}(u,r)\). This is the end of the proof for lemma \ref{Prop2}.

One of the consequence of lemma \ref{Prop2} is the norm of \(\mathcal{F}\) in \(X\) is bounded from above by
\begin{equation}
    \|\mathcal{F}\|_X\leq K_2\frac{d+x^3+x^4+x^5+x^7+x^9}{k^3}\exp\left(K_1(x^2+x^4)\right),
\end{equation}
which implies that finiteness of both \(\|h\|_X\) and \(\|h_0\|_{X_0}\) lead to a bounded \(\|\mathcal{F}\|_X\). Since the map \(\mathcal{F}\) is bounded in \(X\), we proceed to show that \(\mathcal{F}\) is a map from a ball  to itself by introducing the following function
\begin{equation}\label{L1}
    \mathcal{L}_1(x)\equiv\frac{xk^3}{K_2}\exp\left(-K_1(x^2+x^4)\right)-(x^3+x^4+x^5+x^7+x^9) .
\end{equation}
The lemma is   
\begin{lemma}\label{Prop3} Let \(X\) and \(X_0\) are function spaces defined in \eqref{spaceX} and \eqref{spaceX0}, respectively.
Let \(\alpha\) be the effective coupling constant, \(x=\|h\|_X\) and \(x_0\) be the root of \(k(x)=0\) where \(k(x)\) is defined in \eqref{k}. If \(d=\|h_0\|_{X_0}\) as the initial data of dynamical equation \eqref{DynEqOri} satisfies \(d<\mathcal{L}_1(x)\), where \(\mathcal{L}_1(x)\) is defined in \eqref{L1}, then there exist \(x_1\in(n\pi,x_0]\) and \(\alpha\) which takes values from the interval \eqref{intAlpha} such that for every \(x\in(n\pi,x_1)\), we have \(\|\mathcal{F}\|_X<x\) with \(\mathcal{F}:h\rightarrow \mathcal{F}(h)\) defined in \eqref{F}. 
\end{lemma}

\textbf{Proof:} It is straightforward to check that we can always choose \(\alpha\) small enough such that \(\mathcal{L}_1(n\pi)>0\) and \(\mathcal{L}_1'(n\pi)>0\). Since \(\mathcal{L}_1(x\rightarrow \infty)=-\infty\) then there exist a finite \(x_1\geq n\pi\) where \(\mathcal{L}_1(x)\) is monotonically increasing for \(x\in(n\pi,x_1)\). Since we can always, again, choose \(\alpha\) to be small enough such that \(x_0>n\pi\), then we can always choose to take \(x_1\) inside \([0,x_0]\). Hence, if \(d<\mathcal{L}_1(x)\) for all \(x\in(n\pi,x_1)\), then \(\|\mathcal{F}\|_X<x\).

The role of lemma \ref{Prop3} is obvious as the first necessary condition for \(\mathcal{F}\) to have a fixed point as a contractive mapping. Suppose that we have an open ball \(\mathcal{B}(n\pi,x)=\{h\in X | \|h\|_X \in (n\pi,x)\}\). The fact that \(\|\mathcal{F}(h)\|_X<\|h\|_X\) is always true for all \(x\in(0,x_1)\) implies that \(\mathcal{F}:\mathcal{B}(n\pi,x)\rightarrow \mathcal{B}(n\pi,x)\). Thus, the first necessary condition is satisfied. The role of \(x_1\) itself is to be the maximum possible value of \(x\) (i.e. the maximum radius of the ball), hence, the contractive property of \(\mathcal{F}\) must be proven for \(\|h\|_X\leq x_1\) in which \(\mathcal{F}:\mathcal{B}(n\pi,x)\rightarrow \mathcal{B}(n\pi,x)\) is guaranteed. 

The condition \(d<\mathcal{L}_1(x)\) implies that the initial condition, \(h_0\), is not arbitrary and it must be small enough in order to guarantee that the necessary condition for \(\mathcal{F}\) is satisfied. This came from the fact that even though we are free to choose \(K_1\) and \(K_2\), both of them are bounded from below. Thus, the highest upper bound for \(d\) actually depends on the effective gravitational coupling constant, \(\alpha\), since the lower bound of \(K_1\) and \(K_2\) depend on \(\alpha\) and the bound for \(d\) just get smaller for a higher value of coupling constant. We would like to remark that not all values of \(\alpha\) in the interval \eqref{intAlpha}. This is coming from the fact that the conditions from Lemma \ref{Prop3} and Lemma \ref{GlobalLowBound} give a stronger constraint for \(\alpha\) because we need \(\mathcal{L}_1(x)>n\pi\) for \(x\in(n\pi,x_1)\) but this constraint depends on \(n\) and should be calculated numerically for each different \(n\), but it is obvious that the upper bound for \(\alpha\) corresponding to this condition is finite and non-zero. 

Now, the remaining condition needed to be satisfied is that the upper bound of \(\mathcal{L}_1(x)\) cannot be less than \(n\pi\). We can show it by proving that there exists \(\alpha\) (which should take values in the intervals described in the previous paragraph) such that \(\mathcal{L}_1(n\pi)>n\pi\). Let \(\alpha_m\) be the value of coupling constant such that \(\mathcal{L}_{1}(n\pi)=n\pi\). If \(\alpha=0\) then \(\mathcal{L}_{1}(n\pi)\) is unbounded (since \(K_2\geq0\)). Thus, for \(0\leq\alpha<\alpha_m\), \(\mathcal{L}_{1}(n\pi)\) is strictly higher than \(n\pi\), implying that there exist \(d\) which takes values between \(\mathcal{L}_{1}(n\pi)\) and \(n\pi\).  This completes the proof of Lemma \ref{Prop3}.

Since the proof for the first property of the existence of unique solutions has been established, we proceed by showing that \(\mathcal{F}\) contracts in \(Y\).
\begin{lemma}
Suppose that \(h_1, h_2\in X\) solves \eqref{DynEqOri} and both share exactly the same initial data \(h_0\in X_0\), where \(X\) and \(X_0\) defined in \eqref{spaceX} and \eqref{spaceX0}, respectively. Let \(Y\) be a space defined in \eqref{spaceY} and \(\mathcal{F}\) be a map defined in \eqref{F}. There exist \(\lambda\in(0,1)\) such that 
\begin{equation}
    \|\mathcal{F}(h_2)-\mathcal{F}(h_1)\|_Y\leq \lambda \|h_2-h_1\|_Y.
\end{equation}
\end{lemma}
\textbf{Proof:} By following the same method demonstrated in \cite{Chae:2001ec} and in the previous discussion on the local existence, we define \(\Delta \equiv \mathcal{F}(h_2)-\mathcal{F}(h_1)\) and take \(x\) to be greater than \(\max \{\|h_1\|_X,\|h_2\|_X\}\). Then, from \eqref{DynEqOri} we have the integro-differential equation for \(\Delta\) along the curve \(\chi\) that is given by \eqref{dynEqDelTilde}. By assuming that \(h_1\) and \(h_2\) have exactly the same initial data, the solution of \eqref{dynEqDelTilde} for this setup is given by
\begin{equation}
    \Delta=\int_0^{u'} \exp\left(\int_u^{u'} P_2 d\Tilde{u}\right)\left[(P_2-P_1)\mathcal{F}(h_1)+\frac{\Tilde{g}_2-\Tilde{g}_1}{2}\mathcal{G}(h_1)+(Q_2-Q_1)\right]du.
\end{equation}
Firstly, we estimate the first term, \(\int_0^{u'}(P_2-P_1)\mathcal{F}(h_1)du\), as follows: the integrand can be estimated as
\begin{eqnarray}
    (P_2-P_1)\mathcal{F}(h_1)&\leq& \left[\left|\frac{g_2-g_1-\Tilde{g}_2+\Tilde{g}_1}{2r}\right|+\left|\frac{\alpha}{2} r\left(2\frac{\Bar{h}^2_2-\Bar{h}^2_1}{r^2}+\frac{\Bar{h}^4_2-\Bar{h}^4_1}{r^4}\right)\right|\right.\nonumber\\&&+\left.\frac{d}{du'}\left|2\frac{\Bar{h}^2_2-\Bar{h}^2_1}{r^2}\right|\right]\mathcal{F}(h_1).
\end{eqnarray}
Let \(y=\|h_2-h_1\|_Y\), then we have
\begin{equation}
    |\Bar{h}_2-\Bar{h}_1|\leq \frac{y}{(1+u)(1+r+u)},
\end{equation}
which leads to
\begin{eqnarray}
    |\Bar{h}^2_2-\Bar{h}^2_1|&\leq&\frac{2xy}{(1+u)^2(1+r+u)^2}\\
    |\Bar{h}^4_2-\Bar{h}^4_1|&\leq&\frac{4x^3y}{(1+u)^4(1+r+u)^4}\\
    |g_2-g_1-\Tilde{g}_2+\Tilde{g}_1|&\leq&\alpha \frac{2(x+x^3)yr^2}{3(1+u)^3(1+r+u)^3}.
\end{eqnarray}
Thus, we have
\begin{eqnarray}
    |(P_2-P_1)\mathcal{F}(h_1)|&\leq&\left[\alpha \frac{(x+x^3)yr}{3(1+u)^3(1+r+u)^3}+4\alpha \frac{(x+x^3)y}{r(1+u)^2(1+r+u)^2}\right.\nonumber\\&&+\left.\frac{d}{du'}\left(4\frac{xy}{r^2(1+u)^2(1+r+u)^2}\right)\right]K_1\frac{d+x^3+x^5}{k^2(1+r+u)^2}\exp(K_1 (x^2+x^4)).\nonumber\\
\end{eqnarray}
As such, the estimate for the first term is given by
\begin{eqnarray}
    \int_0^{u'}(P_2-P_1)\mathcal{F}(h_1)du&\leq&4\alpha \frac{K_1}{k^2}(x+x^3)(d+x^3+x^5)y\nonumber\\&&\exp(K_1 (x^2+x^4))\int_0^{u'}\frac{du}{(1+u)^2(1+r+u)^4}\nonumber\\
    &&+4\frac{K_1}{k^2}\frac{x(d+x^3+x^5)y}{r^2(1+u')^2(1+r'+u')^2}\exp(K_1 (x^2+x^4))\nonumber\\
    &\leq& 16\alpha K_1 \frac{(x+x^3)(d+x^3+x^5)y}{k^4(1+r'+u')^2}\exp(K_1 (x^2+x^4)).
\end{eqnarray}
Now, for the second term, \(\int_0^{u'}\frac{\Tilde{g}_2-\Tilde{g}_1}{2}\mathcal{G}(h_1)du\), we need the following relation 
\begin{eqnarray}
    |\Tilde{g}_2-\Tilde{g}_1|&\leq& \left|\frac{\alpha}{r}\int_0^rs^2\left(2\frac{(h_2-\Bar{h}_2)^2-(h_1-\Bar{h}_1)^2}{s^2}+\frac{(h_2-\Bar{h}_2)^4-(h_1-\Bar{h}_1)^4}{s^4}\right)ds\right|\nonumber\\
    &&+|\Bar{g}_2-\Bar{g}_1|\nonumber\\
    &\leq& \frac{\alpha}{r}\int_0^rs^2\left(2\frac{xy}{2(1+u)^2(1+r+u)^4}+\frac{x^3y}{4(1+u)^4(1+r+u)^8}\right)ds\nonumber\\
    &&+|\Bar{g}_2-\Bar{g}_1|
\end{eqnarray}
which is deduced from equation \eqref{gTildeOri}. This inequality requires estimate of \(|\Bar{g}_2-\Bar{g}_1|\) which can be deduced from \eqref{gOri}. To do so, consider the following relation
\begin{eqnarray}
    |g_2-g_1|&\leq& |h_2-h_1| \left|\frac{\partial g}{\partial h}\right|_{\max(h_2,h_1)}\nonumber\\
    &\leq&\alpha y\left|\int_r^\infty s\frac{x^2}{8(1+u)^2(1+s+u)^4}\left(1+\frac{x^2}{2(1+u)^2(1+s+u)^4}\right)ds\right|\nonumber\\
    &\leq&\alpha  \frac{(x^2+x^4)y}{16(1+u)^2(1+r+u)^2},
\end{eqnarray}
that will be used to estimate \(|\Bar{g}_2-\Bar{g}_1|\) as follows,
\begin{eqnarray}
    |\Bar{g}_2-\Bar{g}_1|&=&\left|\frac{1}{r}\int_0^r(g_2-g_1)ds\right|\leq \frac{1}{r}\int_0^r|g_2-g_1|ds\nonumber\\
    &\leq& \alpha  \frac{(x^2+x^4)y}{16(1+u)^3(1+r+u)}.
\end{eqnarray}
Now that we have the estimate of \(|\Bar{g}_2-\Bar{g}_1|\), we can proceed to estimate the second term as follows
\begin{eqnarray}
    \int_0^{u'}\frac{\Tilde{g}_2-\Tilde{g}_1}{2}\mathcal{G}(h_1)du &\leq& \int_0^{u'} \alpha  \frac{(x^2+x^4)y}{32(1+u)^3(1+s+u)}\nonumber\\&&\left(K_2\frac{d+x^3+x^4+x^5+x^7+x^9}{k^3(1+r'+u')^3}\exp\left(K_1(x^2+x^4)\right)\right) du\nonumber\\
    &\leq& \frac{\alpha}{2}  K_2 \frac{(x^2+x^4)(d+x^3+x^4+x^5+x^7+x^9)}{k^5(1+r'+u')^2}\exp\left(K_1(x^2+x^4)\right).\nonumber\\
\end{eqnarray}
The last term (third) to be estimated is \(\int_0^{u'}\left(Q_2-Q_1\right)du\). First consider the \(|Q_2-Q_1|\) that satisfies
\begin{eqnarray}
    |Q_2-Q_1|&\leq&\left|\frac{\Bar{h}_2(g_2-\Tilde{g}_2)-\Bar{h}_1(g_1-\Tilde{g}_1)}{2r}\right|+\left|\frac{\alpha}{2}  r\left(3\frac{\Bar{h}^3_2-\Bar{h}^3_1}{r^2}+\frac{\Bar{h}^5_2-\Bar{h}^5_1}{r^4}\right)\right|\nonumber\\&&+\frac{d}{du'}\left|2\frac{\Bar{h}^3_2-\Bar{h}^3_1}{r^2}\right|\nonumber\\
    &\leq& 4\alpha  \frac{(x^2+x^4)y}{(1+u)^3(1+r+u)^3}+\frac{d}{du'}\left|4\frac{x^2y}{(1+u)^3(1+r+u)^5}\right|.
\end{eqnarray}
As such, the third term can be estimated as
\begin{eqnarray}
    \int_0^{u'}\left(Q_2-Q_1\right)du&\leq& 4\alpha (x^2+x^4)y\int_0^{u'}\frac{du}{(1+u)^3(1+r+u)^3}+4\frac{x^2y}{(1+u')^3(1+r'+u')^5}\nonumber\\
    &\leq& 8\alpha \frac{(x^2+x^4)y}{k^2(1+r'+u')^2}.
\end{eqnarray}
By combining all of the three estimates of the three terms we have a decay estimate for \(\Delta\), namely
\begin{equation}
    |\Delta(u',r')|\leq \Tilde{K}_2 \frac{(\sigma_1(x)+\sigma_2(x)+x^2+x^4)y}{k^5(1+r'+u')^2}\exp\left(\Tilde{K}_1(x^2+x^4)\right),
\end{equation}
where \(\Tilde{K}_1\geq 8\alpha\), \(\Tilde{K}_2\geq7\alpha^3 \), and the two functions, \( \sigma_1(x)\) and \(\sigma_2(x)\) is given by
\begin{eqnarray}
    \sigma_1(x)&\equiv&(x+x^3)(d+x^3+x^5)\\
    \sigma_2(x)&\equiv&(x^2+x^4)(d+x^3+x^4+x^5+x^7+x^9).
\end{eqnarray}
Since the decay estimate of \(\Delta\) goes like \((1+r'+u')^{-2}\), we can conclude that the norm of \(\Delta\) in \(Y\) satisfies
\begin{equation}
    \|\Delta\|_Y\leq\mathcal{L}_2(x) y,
\end{equation}
where \(\mathcal{L}_2(x)\) is defined as 
\begin{equation}
    \mathcal{L}_2(x)\equiv \frac{\Tilde{K}_2}{k^5}(\sigma_1(x)+\sigma_2(x)+x^2+x^4)\exp\left(\Tilde{K}_1(x^2+x^4)\right).
\end{equation}
The function \(\mathcal{L}_2(x)\) is monotonically increasing for all \(x\in[n\pi,x_1]\) and \(\mathcal{L}_2(n\pi)>0\). Thus, there exist \(x_2\in (n\pi,x_1]\) and \(\alpha\) which falls in the interval \eqref{intAlpha} such that \(\mathcal{L}_2(x)<1\) for all \(x\in (n\pi,x_2)\). Furthermore, the smaller value of \(\alpha\) leads to a smaller \(\mathcal{L}_2(x)\) as well. By taking \(\lambda=\mathcal{L}_2(x)\) for \(\|h\|_X<x_2\), we prove that \(\mathcal{F}\) contracts in \(Y\). This completes our proof of the global existence of the Einstein-Skyrme system for small initial data.
\section{Singularity Formations}\label{Sect4}
From our previous discussion, we have shown that the existence of global smooth solutions can be guaranteed for only small initial data. Thus, there are still large varieties of possible regular initial conditions that could lead to singularities at a finite time. The reason why we are interested in regular initial conditions is that we can model a gravitational collapse of a Skyrme star since we already know that in the static limit, we have a lot of varieties of Skyrme black holes and even a naked singularity inside a star solution \cite{Adam:2016vzf,Gunara:2018lma,Gudnason:2016kuu, Gunara:2018lma}. 

It is known that the Einstein-Klein-Gordon system in which the scalar field has only kinetic term does allow a singularity in finite time for regular initial conditions which leads to the formation of naked singularities from a scalar star \cite{Christodoulou:1994hg}. This singularity turns out to be unstable, thus, it is not physically possible to be observed and cannot be a counter-example of the cosmic censorship conjecture \cite{Christodoulou:1999math}. Since the non-topological sector (\(B=0,\) or \(A_3=A_4=0\)) of the spherically symmetric Skyrme model we consider in this work coincides with the Einstein-Klein-Gordon system, it is interesting to study the behaviour of the topological sector (\(B\neq 0\) or \(|A_3|=|A_4|=1\)) and its singularity developments. 

The main idea to prove that singularity might develop within this system is by looking at the value of the Skyrmion profile function, \(\xi\), at the origin, \(r=0\). Since, the Skyrme field \(\phi(\Vec{x})\) within this choice of ansatz, given in \eqref{ansatzPhi}, must be unique for every \(\Vec{x}\), then the value of Skyrmion profile function cannot be arbitrary at the origin. It has been proven that in the static limit \(\xi\) must be a multiple of \(\pi\) for non-gravitating cases (flat spacetime background) \cite{Skyrme:1961vq,Skyrme:1962vh}. Furthermore, for a static Skyrme blackhole solution where we can only determine the possible values of \(\xi\) at the blackhole horizon, it is hinted that the unique property of \(\phi\) at the origin, found in the non-gravitating case, cannot be satisfied \cite{Adam:2016vzf,Gudnason:2016kuu}. Thus, finding such indications is sufficient to show that this Einstein-Skyrme system develops singularities. In order to do this, we consider the self-similar solutions that are known to produce singularity at spacetime origin for some known scalar field theories, for example, the Einstein-Klein-Gordon model in Bondi coordinate \cite{Christodoulou:1994hg} and \(SU(2)\) sigma model in flat spacetime \cite{Shatah1988WeakSA}. We expect that there exist a family of self-similar solutions within the Einstein-Skyrme system which develops singularity in finite time.

One of the main problems in analyzing the self-similar solutions of the Einstein-Skyrme system is that the two terms in the Skyrme model Lagrangian scales differently under global spatial scaling \(\Vec{x}\rightarrow\mu\Vec{x}\). In \(3+1\) dimensional spacetime the quadratic term scales like \(\mu\) and the quartic term, scales like \(\mu^{-1}\). Thus, the resulting dynamical equation is not scale invariant. This fact is well represented in \eqref{DynEqOri} which cannot be reduced into self-similar form due to the difference in scaling behaviour of each term. 

The approach we introduce here to circumvent this problem is to consider the singularity development of each submodel containing only one of the terms from the dynamical equation. This can be achieved by taking \(C_2=0\) for the quadratic submodel, and \(C_1=0\) for the quartic submodel. In this approach, the resulting terms in dynamical equations of each submodel transforms similarly under coordinate scaling transformation. Thus, we can recast them into self-similar forms.
\subsection{Singularities in Quadratic Submodel}
The submodel we consider here is actually a sigma model possessing the first invariant of strain tensor induced by the field \(\phi:\mathcal{M}^{1,3}\rightarrow S^3\) with spherically symmetric ansatz \cite{manton1987}. The dynamical equation of Skyrmion field in this submodel is given by
\begin{equation}\label{DynEqQuadratic}
    r e^{-2 \sigma} \left(\xi_{,r} \left(-r \rho_{,r}+r \sigma_{,r}-2\right)+2 \left(\xi_{,u}+r \xi_{,ur}\right) e^{\sigma-\rho}-r \xi_{,rr}\right)+\sin (2 \xi )=0,
\end{equation}
with Einstein's Equations governing the dynamics of \(\rho\) and \(\sigma\), given by
\begin{eqnarray}\label{EinsteinQuadratic1}
    r\left(\rho_{,r}+\sigma_{,r}\right)&=&\alpha r^2\xi_{,r}^2~,\\
    e^{2\sigma}-1-r\left(\rho_{,r}-\sigma_{,r}\right)&=&2\alpha e^{2\sigma}\sin^2\xi.\label{EinsteinQuadratic2}
\end{eqnarray}
We can see that the system of equations is now scale invariant. To recast these equations into a self-similar form we introduce \(t=\frac{-2r}{u}\) such that region \(t>0\) lies in the past of apparent horizon \cite{Christodoulou:1994hg}. It is also advantageous to recast \(\{\rho,\sigma\}\) into \(\{g,\tilde{g}\}\) as defined in (\ref{defg},\ref{defTildeg}) such that we can get rid of any possibility of dealing with complex functions behind the horizon by imposing both \(g\) and \(\tilde{g}\) to be at least \(C^1\)-real functions of \(t\). The system of self-similar equations is given by
\begin{eqnarray}\label{SelSimSigma1}
    \left[t^3-t^2\tilde{g}\right]\xi''+\left[2t^2-(2t\tilde{g}+t^2\tilde{g}')\right]\xi'+g\sin\left(2\xi\right)&=&0,\\\label{SelSimSigma2}
    g'-\alpha g t\xi'{}^2&=&0,\\\label{SelSimSigma3}
    \tilde{g}'-\frac{g-\tilde{g}}{t}+2\alpha  g\frac{\sin^2\xi}{t}&=&0.
\end{eqnarray}
From this point, we assume that the self-similar solution \(\xi(t)\) is at least a \(C^2\)-real function.
Let us define the first derivative of \(\xi\) to be \(\psi\equiv\xi'\) and a new vector \(\textbf{w}\equiv \begin{bmatrix}\xi\\\psi\end{bmatrix}\), \(\textbf{w}\in U\subset \mathbb{R}^2\), such that the following lemma holds in closed interval \(\mathcal{I}=[t,t+\epsilon]\), \(\epsilon>0\)
\begin{lemma}\label{LocLipMetric}
The metric functions \(g(\textbf{w},t)\) and \(\tilde{g}(\textbf{w},t)\), that satisfies equations \eqref{SelSimSigma2} and \eqref{SelSimSigma3}, are locally Lipschitz with respect to \(\textbf{w}\).
\end{lemma}
\textbf{Proof: }The Einstein equations \eqref{SelSimSigma2}
 and \eqref{SelSimSigma3} can be rewritten into integral form as follows
\begin{eqnarray}\label{Eqg}
    \ln g&=&\alpha\int t'\xi'{}^2~dt',\\
    \label{EqTildeg}t\tilde{g}&=&\int g\left[1-2\alpha \sin^2\xi\right]~dt'.
\end{eqnarray}
We can see that \eqref{Eqg} implies \(g>0\). Now, consider a solution for \(g\) and a solution for \(\tilde{g}\) in the closed interval \(\mathcal{I}\). Integrating \eqref{Eqg} and \eqref{EqTildeg} in the whole domain gives us the relation
\begin{eqnarray}
    |g(t+\epsilon)|_{U}&\leq& |g(t)|_U\exp \left[C|\alpha t \psi^2|\right],\\
     |\tilde{g}(t+\epsilon)|_{U}&\leq& |\tilde{g}(t)|_U+C\left|\frac{g}{t}\left[1-2\alpha \sin^2\xi\right]\right|
\end{eqnarray}
with \(C>\epsilon\), implying that both \(g\) and \(\tilde{g}\) are bounded. Then for \(\textbf{w}_1,\textbf{w}_2\in U\) we have
\begin{eqnarray}
    |g(\textbf{w}_1,t)-g(\textbf{w}_2,t)|_{U}&\leq& \left|\exp \left[C|\alpha t \psi_1^2|\right]-\exp \left[C|\alpha t \psi_2^2|\right]\right|,\\
    |\tilde{g}(\textbf{w}_1,t)-\tilde{g}(\textbf{w}_2,t)|_{U}&\leq& C g(\textbf{w},t)\left|\frac{2\alpha }{t}\left(\sin^2\xi_1-\sin^2\xi_2\right)\right|
\end{eqnarray}
Where \(\textbf{w}\in \{\textbf{w}_1,\textbf{w}_2\}\) which maximizes \(g(\textbf{w},t)\). Since \(g\) and \(\tilde{g}\) are bounded and are smooth functions of \(\textbf{w}\), then we can apply a local property
\begin{equation}\label{LocProp}
    |F(f_1)-F(f_2)|\leq \sup_{s\in[0,1]}\left[F'(f_1+s(f_2-f_1))\right](f_1-f_2),
\end{equation}
to see that \(g\) and \(\tilde{g}\) are locally Lipschitz with Lipschitz constants that are functions of \(\textbf{w}\in U\),
\begin{eqnarray}
    |g(\textbf{w}_1,t)-g(\textbf{w}_2,t)|_{U}&\leq& C_g(\|\textbf{w}_1\|,\|\textbf{w}_2\|)\|\textbf{w}_1-\textbf{w}_2\|,
    \\
    |\tilde{g}(\textbf{w}_1,t)-\tilde{g}(\textbf{w}_2,t)|_{U}&\leq& C_{\tilde{g}}(\|\textbf{w}_1\|,\|\textbf{w}_2\|)\|\textbf{w}_1-\textbf{w}_2\|
\end{eqnarray}
Both \(C_g\) and \(C_{\tilde{g}}\) are finite real functions of the norms of \(\textbf{w}_1\) and \(\textbf{w}_2\).

From here, to check the existence of unique solutions, we rewrite \eqref{SelSimSigma1} into 
\begin{equation}\label{VecFromEqXi}
    \frac{d}{dt}\textbf{w}=\mathcal{J}_1(\textbf{w},t),
\end{equation}
where we have defined
\begin{equation}
    \mathcal{J}_1(\textbf{w},t)\equiv\begin{bmatrix}\psi\\
    J_1\end{bmatrix},
\end{equation}
with
\begin{equation}
    J_1=\left(\frac{g+\tilde{g}-2t}{t(t-\tilde{g})}-2\alpha g \frac{\sin^2\xi}{t(t-\tilde{g})}\right)\psi-\left(\frac{g}{t^2(t-\tilde{g})}\right)\sin(2\xi).
\end{equation}
Note that the resulting system of equations is for \(\xi\) that is constrained by equations of metric function from Lemma \eqref{LocLipMetric}. We can now state the local property of \eqref{VecFromEqXi} as follows
\begin{lemma}\label{LipLocJ1}
The operator \( \mathcal{J}_1(\textbf{w},t)\) in \eqref{VecFromEqXi} is locally Lipschitz with respect to \(\textbf{w}\).
\end{lemma}
\textbf{Proof: }From the definition of \(J_1\) we know that for every \(\textbf{w}\in U\) we can estimate
\begin{equation}\label{Jest}
    |J_1|_{U}\leq \left|\frac{g+\tilde{g}}{t(t-\tilde{g})}\psi\right|+2\alpha g\left| \frac{\sin^2\xi}{t(t-\tilde{g})}\psi\right|+\left|\frac{g}{t^2(t-\tilde{g})}\sin(2\xi)\right|.
\end{equation}
Because \(\xi(t)\) belongs to the class of \(C^2\)-real functions, then it is bounded on any closed \(\mathcal{I}\). This implies that \(|\mathcal{J}_1(\textbf{w},t)|\) is bounded in \(U\). Then, for \(\textbf{w}_1,\textbf{w}_2\in U\) we have
\begin{eqnarray}
    \|\mathcal{J}_1(\textbf{w}_1,t)-\mathcal{J}_1(\textbf{w}_2,t)\|_U&\leq& \left|1+\frac{g+\tilde{g}}{t(t-\tilde{g})}\right||\psi_1-\psi_2|\nonumber\\
    &&+\left|\frac{2\alpha g}{t(t-\tilde{g})}\right||\psi_1\sin^2\xi_1-\psi_2\sin^2\xi_2|\\
    &&+\left|\frac{g}{t^2(t-\tilde{g})}\right||\sin(2\xi_1)-\sin(2\xi_2)|\nonumber,
\end{eqnarray}
with \(g=g(\textbf{w},t)\) and \(\tilde{g}=\tilde{g}(\textbf{w},t)\) where \(\textbf{w}\in \{\textbf{w}_1,\textbf{w}_2\}\) which maximizes \(g+\tilde{g}\). From Lemma \eqref{LocLipMetric} we know that the metric functions are locally Lipschitz. Thus, by employing the property \eqref{LocProp} for operator \(\mathcal{J}_1(\textbf{w},t)\) we have
\begin{equation}
    \|\mathcal{J}_1(\textbf{w}_1,t)-\mathcal{J}_1(\textbf{w}_2,t)\|_U\leq C_{\mathcal{J}_1}(\|\textbf{w}_1\|,\|\textbf{w}_2\|)\|\textbf{w}_1-\textbf{w}_2\|.
\end{equation}
This inequality shows that \(\mathcal{J}_1(\textbf{w},t)\) is locally Lipschitz with a Lipschitz constant that is a function of \(\textbf{w}\).
\begin{remark}\label{remarkMetric}
We have two conditions that is to be satisfied in order to extend the validity of Lemma \eqref{LipLocJ1} for intervals that contains \(t=0\), namely
\begin{itemize}
    \item \(\xi(t=0)=n\pi\), with \(n\in\mathbb{Z}\), and
    \item  \(\lim_{t\rightarrow0}\psi=0\) or 
    \(\lim_{t\rightarrow0}(g+\tilde{g})=0\).
\end{itemize}
The condition \(\lim_{t\rightarrow0}\psi=0\) gives vacuum solutions and the other condition, \(\lim_{t\rightarrow0}(g+\tilde{g})=0\), implies changes of sign in metric tensor components which indicate the presence of horizon near the singularity.
\end{remark}
The explanation for two conditions given above is straightforward. Firstly, if the first condition is satisfied, then \(|\psi(t)-\psi(0) | =|\psi(t)|\leq C|t|\) because \(\xi(t)\) belongs to \(C^2\)-real functions. Then, secondly, if the second condition holds true and the condition for \(\xi(t=0)\) is satisfied, then the first term  of \eqref{Jest} is bounded, and the rest of the terms containing \(\sin\xi\) is bounded. Note that we only need one of the optional conditions to ensure the boundedness of \(|J_1|_U\). As such, if one of them is satisfied then Lemma \eqref{LipLocJ1} can be extended to intervals that contains \(t=0\). Since vacuum solutions will not develop any singularities, we are now assume that the second optional condition holds and \(\psi(0)=\psi_0\neq0\).

Now, consider the solution of \eqref{VecFromEqXi} in \(\mathcal{I}\) that is given by
\begin{equation}
    \textbf{w}(\tilde{t})=\textbf{w}(t)+\int_{t}^{\tilde{t}} \mathcal{J}_1(\textbf{w},t')dt',
\end{equation}
with \(\tilde{t}\in \mathcal{I}\). By Introducing a Banach Space 
\begin{equation}
    W\equiv\{\textbf{w}\in C(\mathcal{I},\mathbb{R}^2):\textbf{w}(t)=\textbf{w}_0,\sup_{\tilde{t}\in\mathcal{I}}|\textbf{w}(\tilde{t})|\leq L_0\},
\end{equation}
equipped with norm
\begin{equation}
    \|\textbf{w}\|_{W}=\sup_{\tilde{t}\in\mathcal{I}}|\textbf{w}(\tilde{t})|,
\end{equation}
where \(L_0\) is a positive constant, we define an operator \(\mathcal{K}:W\rightarrow W\),
\begin{equation}\label{OpK}
    \mathcal{K}(\textbf{w}(\tilde{t}))=\textbf{w}_0+\int_{t}^{\tilde{t}} \mathcal{J}_1(\textbf{w},t')dt'.
\end{equation}
As such, Lemma \eqref{LipLocJ1} implies the uniqueness of the solution of the differential equation \eqref{VecFromEqXi}.
\begin{corollary}
The operator \(\mathcal{K}\) defined in \eqref{OpK} is a map from \(W\) to itself and is a contraction mapping for \(w\in W\) that is defined in a closed interval \(\mathcal{I}=[t,t+\epsilon]\), \(t\in\mathbb{R}\), \(\epsilon>0\), and 
\begin{equation}
    \epsilon\leq \min\left[\frac{1}{C_{L_0}},\frac{1}{C_{L_0}L_0+\|\mathcal{J}_1(t)\|}\right].
\end{equation}
Thus, there exists a unique smooth solution \(\textbf{w}\in W\) of differential equation \eqref{VecFromEqXi}.
\end{corollary}
We can construct the maximal solution from the solutions we consider above up to any finite \(t_m>0\) as follows: Suppose that \(\textbf{w}(t)\) be a solution on the interval \([0,t_m]\) with initial values \(\textbf{w}(0)=\begin{bmatrix}n\pi\\\psi_0\end{bmatrix}\), \(n\in\mathbb{Z}\). We can repeat the arguments given above with new initial values on \(\textbf{w}(t-t_k)\) where \(0<t_k<t_m\) and then proceed to employ the uniqueness condition to glue the resulting solutions up to \(t=t_m\). From here, we are ready to state the theorem on the singularity formations of the first submodel in the Einstein-Skyrme system.
\begin{theorem}\label{TheorSing}
There exists smooth initial data for the quadratic submodel of the Einstein-Skyrme system in the Bondi coordinates that develops singularity at the coordinate origin in finite \(u\).
\end{theorem}
\textbf{Proof :} In order to prove that such singularity could develop, we shall use the self-similar solutions above as an example of initial data. The spherically symmetric Einstein-Skyrme system's dynamics in Bondi coordinate is governed by three equations for three dynamical fields, namely equation \eqref{DynEqQuadratic}, \eqref{EinsteinQuadratic1}, and \eqref{EinsteinQuadratic2}. Let \([\xi_0(r),g_0(r),\tilde{g}_0(r)]\) be smooth functions that are the solutions of system of equations \eqref{SelSimSigma1}, \eqref{SelSimSigma2}, and \eqref{SelSimSigma3}, satisfying a set of initial values given by
\begin{equation}
    \xi_0(0)=0,~~~\xi_0'(0)=\psi_0\neq0,~~~g_0(0)+\tilde{g}_0(0)=0,
\end{equation}
where \(\psi_0\) is an arbitrary non-zero real number. Let [\(\xi(u,r),g(u,r),\tilde{g}(u,r)\)] be the solutions of the system \eqref{DynEqQuadratic}, \eqref{EinsteinQuadratic1}, and \eqref{EinsteinQuadratic2}, with initial data at \(u=-2\) given by
\begin{equation}
    \xi(-2,r)=\xi_0(r),~~~ \partial_u\xi(-2,r)=\frac{r}{2}\partial_r\xi_0(r),~~~g(-2,r)=g_0(r),~~~\tilde{g}(-2,r)=\tilde{g}_0(r).
\end{equation}
Then, in the past light-cone extended from \(u=-2\) to the vertex \((u=0,r=0)\), the solution of \(\xi(u,r)\) is given by
\begin{equation}
    \xi(u,r)=\xi_0\left(-\frac{2r}{u}\right),
\end{equation}
hence satisfies 
\begin{equation}
    \partial_r\xi(u,0)=-\frac{2}{u}\partial_r\xi_0(0).
\end{equation}
Because \(\partial_r\xi_0(0)=\psi_0\neq0\) then \(\partial_r\xi(u,0)\rightarrow\infty\) as \(u\rightarrow0\). From Einstein's equation, we can see that the singular point for \(\xi\) is also a singular point for both \(g\) and \(\tilde{g}\), resulting in a singularity for the metric functions as well. We would like to remark that the singularity from the self-similar solution demonstrated above is hidden behind a horizon (see Remark \ref{remarkMetric}), in contrast with the one found from a similar setup for the Einstein-Klein-Gordon system where the singularity is naked (does not located inside the horizon) \cite{Christodoulou:1994hg,Christodoulou:1999math}. Interestingly, the solutions keep the topological charge to be \(B=\pm n\) during the process of singularity formation because \(\lim_{r\rightarrow 0}\xi=n\pi\) for all \(u\in[-2,0)\). The value of the topological charge might change after the singularity is formed, \(u=0\), but the solutions should behave differently and the governing equations (4.1-4.3) cannot be used anymore due to the existence of horizon. This might be related to the fact that the Skyrme black holes usually possess a fractional topological charge \cite{Bizon:1992gb,Droz:1991cx,Shiiki:2005xn}. 

\subsection{Remarks on The Quartic Submodel}
In this subsection, we are going to discuss the singularity formation for the second submodel. As mentioned before, the different scaling behaviour of the quartic (Skyrme) term with the rest of the terms in the field equations raises a problem when we try to introduce the same self-similar form from the previous case for the quadratic submodel. To see this explicitly, consider the following expression of Einstein equations for the quartic submodel
\begin{eqnarray}
    r(\rho_{,r}+\sigma_{,r})&=&2\alpha \xi_{,r}^2 \sin^2\xi,\\
    e^{2\sigma}-1-r(\rho_{,r}-\sigma_{,r})&=&\alpha  e^{2\sigma}\frac{\sin^4\xi}{r^2}.
\end{eqnarray}
We can see that, if we impose the same assumption that \(\sin\xi\) is invariant under rescaling of \(r\) then the right-hand and left-hand side of both equations above scales differently, but this is not the case for the Skyrme field equation that is given by
\begin{eqnarray}
    -2(\cos\xi)_{,ur}+g\sin^2\xi\frac{\cos\xi}{r^2}+\left[\tilde{g}_{,r}(\cos\xi)_{,r}+\tilde{g}(\cos\xi)_{,rr}\right]=0.
\end{eqnarray}
In order to keep the same form of solution as the previous submodel, here we can only consider the limit \(\alpha=0\). Thus, the Einstein equations after substitution of \(g\) and \(\tilde{g}\), and then followed by assumption that both \(g\) and \(\tilde{g}\) are functions of \(t=\frac{-2r}{u}\), are now given by
\begin{equation}
    g'=0,~~~\tilde{g}'=\frac{g-\tilde{g}}{t}.
\end{equation}
The first implication is that \(g\) is a constant that should be chosen as 1 since the metric components are flat at the asymptotic boundary. Then, the second implication of the Einstein equation shown above is that \(\tilde{g}\) takes a solution that blows up at \(t=0\), namely
\begin{equation}
    \tilde{g}=1-\frac{l}{t}.
\end{equation}
with \(l\) arbitrary integration constant. As such, \(r=0\) with arbitrary \(u\) is singular points of the metric functions.

Now, the substitution of this form of the metric function to the field equation of this submodel gives
\begin{eqnarray}
    -\left(v'+tv''\right)+(1-v^2)\frac{v}{t^2}+\left[\frac{l}{t^2}v'+\left(1-\frac{l}{t}\right)v''\right]=0.
\end{eqnarray}
where we have defined a new function \(v(t)\equiv \cos(\xi(t))\) such that \(v\) takes values in \([-1,1]\). We can recast the second-order differential equation of \(v\) into two first-order equation by introducing \(\textbf{V}=\begin{bmatrix}v\\z\end{bmatrix}\) with \(z=v'\), such that the resulting system f differential equation is given by
\begin{equation}\label{VecFromEqV}
    \frac{d}{dt}\textbf{V}=\mathcal{J}_2(\textbf{V},t),
\end{equation}
where we have defined
\begin{equation}
    \mathcal{J}_2(\textbf{w},t)\equiv\begin{bmatrix}v\\
    J_2\end{bmatrix},
\end{equation}
with
\begin{equation}
    J_2=\left(\frac{t^2-l}{t(t-l-t^2)}\right)z-\left(\frac{1}{t(t-l-t^2)}\right)\left(1-v^2\right)v.
\end{equation}
Because \(l\) is arbitrary, we have three possible cases to be considered. The first one is \(l>1/4\) where the possibility of \(J_2\) become unbounded is located only at \(t=0\). The second one is \(l=0\) where \(J_2\) possibly become unbounded at \(t=0\) and \(t=1\). The last one is \(0<l\leq1/4\) where we have three possible singular points for \(J_2\), implying that we might need a strong condition for \(\textbf{V}\) in order to make \(J_2\) bounded for every possible interval. 

The resulting submodel does not belong to the Einstein-Skyrme system anymore since we get rid of the back-reaction between gravity and the Skyrme field by choosing \(\alpha=0\) and further details of this submodel are going to be addressed in future works.

\section{Conclusions and Outlook}
\label{sec:concl} 
We have considered the spherically symmetric Einstein-Skyrme system with ansatz that depends on the radial coordinate, \(r\), and retarded time, \(u\). The resulting system of dynamical equations consists of an integro-differential equation for the Skyrme field profile function constrained by two equations from the Einstein field equations governing the profile of metric functions. Firstly, It is shown that unique local classical solutions exist with bounded time intervals that depend on the norm of the initial data, shown in Theorem \ref{TheoremLocal}. The corresponding solutions cannot be extended into global solutions because the time interval cannot be extended to infinity due to the upper bound which comes from the topology of the Skyrmion. Thus the proof of the existence of unique global solutions requires a different strategy. 

Secondly, we show that in order for the global unique solutions to exist, the initial data and and the coupling constant, \(\alpha\), cannot be arbitrary. The corresponding solutions and their first derivative with respect to \(r\) are bounded above by decaying functions stated in Theorem \ref{TheoremGlobal}. Furthermore, we have shown that there exist some possible configurations of initial data within the Einstein-Skyrme system that produce singularity at the light cone vertex. As expected, this singular solution does not belong to the family of global unique solutions from Theorem \ref{TheoremGlobal}.

These results open up the possibility of studying the dynamical stability of the Einstein-Skyrme system, provided the value of the norm of initial data and the coupling constant are inside the allowed intervals in order to make sure that the classical solutions exist. Unfortunately, if we consider the more general case of unrestricted initial data, then the existence of global classical solutions is not guaranteed, as implied by Theorem \ref{TheoremLocal} and Theorem \ref{TheorSing}. However, it is actually possible to circumvent this problem by introducing the concept of a generalized solution for the Einstein-Klein-Gordon system proposed in \cite{Christodoulou:1986zr}. Some modifications to the concept of generalized solutions mentioned above are expected in order to apply it to the Einstein-Skyrme system since we have internal symmetries and topology from the Skyrme fields. Thus, we are going to address these problems in the future.

\section*{Acknowledgements}
The work in this paper is supported by GTA Research Group ITB, Riset ITB 2022, and PDUPT Kemenristek-ITB 2022. A. N. A. would like to acknowledge the support from the  ICTP through the Associates Programme (2018-2023). B. E. G. would also acknowledge the support from the ICTP through the Associates Programme (2017-2022) and BRIN Visiting Researcher Programme 2023. E. S. F. also would like to acknowledge the support from BRIN through the Research Assistant Programme 2022.


%

\bibliography{referenceER}
\bibliographystyle{unsrt}
\end{document}